\documentclass[journal]{IEEEtran}
\usepackage{cite}
\usepackage{comment}
\usepackage{amsmath,amssymb,amsfonts}
\usepackage{algorithmic}
\usepackage{graphicx}
\usepackage{textcomp}
\usepackage{hyperref}
\usepackage{multirow}
\usepackage{multicol}
\usepackage{makecell}
\usepackage{tabularx}
\usepackage{booktabs}
\usepackage{array}
\usepackage[table]{xcolor}
\usepackage[ruled,vlined]{algorithm2e}
\DeclareRobustCommand{\vincenzo}[1]{\textcolor{black}{#1}}

\newcommand{\revisiontwo}[1]{\textcolor{black}{#1}}
\DeclareRobustCommand{\vincenzo}[1]{\textcolor{black}{#1}}

\begin{document}
\title{Diffusion-Based Quality Control of Medical Image Segmentations across Organs}
%\author{Vincenzo Marciano, Hava Chaptoukaev,} \\
%\author{Virginia Fernandez,
%M. Jorge Cardoso, S\'{e}bastien Ourselin, Michela Antonelli, 
%and Maria A. Zuluaga}
\author{
{\rm Vincenzo Marcian\`o,}
\and
{\rm Hava Chaptoukaev,}
\and
{\rm Virginia Fernandez,}
\and
{\rm M. Jorge Cardoso,}\\
\and
{\rm S\'{e}bastien Ourselin,}
\and
{\rm Michela Antonelli,}
\and
{\rm Maria A. Zuluaga}
}

\maketitle

\begin{abstract}
Medical image segmentation using deep learning (DL) has enabled the development of automated analysis pipelines for large-scale population studies. However, state-of-the-art DL methods are prone to hallucinations, which can result in anatomically implausible segmentations. With manual correction impractical at scale, automated quality control (QC) techniques have to address the challenge. While promising, existing QC methods are organ-specific, limiting their generalizability and usability beyond their original intended task. To overcome this limitation, we propose \textit{no-new Quality Control (nnQC)}, a robust QC framework based on a diffusion-generative paradigm that self-adapts to any input organ dataset. Central to nnQC is a novel \textit{Team of Experts (ToE)} architecture, where two specialized \textit{experts} independently encode 3D spatial awareness, represented by the
relative spatial position of an axial slice, and anatomical information derived from visual
features from the original image. A weighted conditional module dynamically combines the pair of independent embeddings, or \textit{opinions} to condition the sampling mechanism within a diffusion process, enabling the generation of a spatially aware pseudo-ground truth for predicting QC scores. Within its framework, nnQC integrates \textit{fingerprint} adaptation to ensure adaptability across organs, datasets, and imaging modalities.
We evaluated nnQC on seven organs using \vincenzo{fifteen} publicly available datasets. Our results demonstrate that nnQC consistently outperforms state-of-the-art methods across all experiments, including cases where segmentation masks are highly degraded or completely missing, confirming its versatility and effectiveness across different organs.
%The pair of independent embeddings, or \textit{opinions}   an image and its axial relative position, generating} a pair of independent embeddings, or \textit{opinions}. A weighted conditional module combines the opinions to \vincenzo{condition} the sampling mechanism within a diffusion process, enabling the accurate generation of a spatially aware pseudo-ground truth (pGT) for predicting QC scores. 
%\vincenzo{By incorporating extracted dataset information, or \textit{fingerprints}, the model can adapt to any given dataset}. 

%We then evaluated nnQC using twelve publicly available datasets encompassing three imaging modalities. Our results demonstrate that nnQC consistently outperforms state-of-the-art methods across all experiments, including cases where segmentation masks are highly degraded or completely missing, confirming it as a versatile and effective QC solution across different organs.
\end{abstract}


\section{Introduction}
\label{sec:introduction}
\IEEEPARstart{A}{dvances} in deep learning (DL) have demonstrated unprecedented capabilities in automating and expediting medical image segmentation~\cite{zhou2021review}.
Despite their high accuracy, DL techniques can still predict anatomically implausible segmentations~\cite{bernard2018deep}. As a result, their translation to real-world clinical applications requires visual quality control (QC). Visual QC process involves inspecting each segmented image for spurious results, followed by manual correction or discarding, which is unfeasible at scale~\cite{Robinson2019Automated,doi:10.1073/pnas.2216399120}. 

Automated QC techniques have emerged as a mechanism to bypass visual QC of predicted segmentations~\cite{kohlberger2012evaluating,valindria2017reverse}. These methods involve the definition of a normative model of high-quality segmentations, which is then used to infer a qualitative~\cite{kohlberger2012evaluating} or quantitative~\cite{doi:10.1073/pnas.2216399120,valindria2017reverse,kalkhof2023m3d,qiu2023qcresunet,lin2022novel,robinson2018realtime,audelan2019unsupervised,fournel2021medical,galati2021efficient,specktor2025segqc,wang2020deep,liu2019alarm} score reflecting the quality of a given predicted image segmentation. 
However, while medical image segmentation frameworks are increasingly general~\cite{MedSAM} or easy to adapt and apply~\cite{isensee2021nnu} across diverse modalities and organs, automated QC methods have not followed the same generalization trend. Current automatic QC methods are limited by their metric-specific~\cite{robinson2018realtime,kohlberger2012evaluating,fournel2021medical} or organ-specific design~\cite{galati2021efficient,lin2022novel,wang2020deep}, restricting their use across anatomical structures and imaging modalities and making their seamless use across applications difficult~\cite{zhou2021review}. As a result, the efficiency and scalability achieved with general-purpose segmentation models are often undermined by the need to design and adapt dedicated QC pipelines for each new organ or application. Enabling large-scale population studies requires robust and self-adapting QC frameworks capable of jointly working with state-of-the-art segmentation methods~\cite{isensee2021nnu} and able to assess segmentations of varying quality and degradation from different organs.

%Current automatic QC methods are limited by their organ-specific design, restricting their use across anatomical structures and imaging modalities~\cite{audelan2019unsupervised,fournel2021medical,galati2021efficient,karani2021test,liu2019alarm,wang2020deep}.

%~\cite{robinson2018realtime,fournel2021medical}. %
% These methods typically involve training a separate model to evaluate segmentation plausibility, using either handcrafted features~\cite{kohlberger2012evaluating}, segmentation uncertainty \cite{qiu2023qcresunet}, or auxiliary predictions such as structural priors or learned shape representations \cite{audelan2019unsupervised}. Some methods also employ supervised approaches, where the QC model learns to regress a quality score (e.g., Dice or Hausdorff distance) from image-segmentation pairs~\cite{fournel2021medical,Robinson2019Automated}, whereas others embed the QC module in the proprietary segmentation model to assess the quality of the self-produced segmentations in an end-to-end fashion \cite{doi:10.1073/pnas.2216399120,kalkhof2023m3d}.

In this work, we introduce \textit{no-new Quality Control} (nnQC), a self-adapting, metric- and model-agnostic framework designed for robust QC of medical image segmentations. nnQC is built upon a Latent Diffusion Model (LDM) backbone and is designed to handle diverse segmentation qualities while remaining adaptable across a wide range of organs, datasets, and imaging modalities. nnQC follows a state-of-the-art 2D reconstruction-based QC approach that generates a pseudo-ground-truth (pGT) mask associated with the predicted segmentation, enabling the estimation of any segmentation quality scores. It introduces a novel sampling strategy, denoted the Team of Experts (ToE), designed to inject 3D contextual information into the pGT reconstruction process. This strategy dynamically balances two independent sources of anatomical insight - referred to as \textit{opinions} - obtained from two separate \textit{experts} that encode anatomical information from the input image and spatial location derived from the segmentation mask. These \textit{opinions} are fused into a conditional vector that guides the LDM's sampling process, enabling anatomically informed generation of the pGT.

Furthermore, inspired by the nnU-Net framework~\cite{isensee2021nnu}, nnQC incorporates the extraction of dataset-specific attributes, or \textit{fingerprints}, enabling seamless self-adaptation across different organs, datasets, and imaging modalities. \vincenzo{As such, the name \textit{nnQC} (no-new QC) reflects the fact that the framework does not require designing a new QC model when adapting to different organs, modalities, or datasets.} We perform an extensive validation of nnQC across \vincenzo{15} diverse scenarios, covering \vincenzo{15} datasets, four imaging modalities or techniques, and seven different organs, demonstrating its generalizability and robustness. To promote reproducibility and encourage broader adoption, we publicly release our open-source code and pre-trained model weights at {\href{https://github.com/robustml-eurecom/nnQC}{github.com/robustml-eurecom/nnQC}}.

\section{Related Works}
%\revision{This section needs to be expanded}

\subsection{Automatic QC}
Automatic QC methods can be categorized into three main classes: embedded, semi-detached, and independent. Embedded QC methods are integrated within the segmentation model itself, allowing the model to self-evaluate its predicted output~\cite{doi:10.1073/pnas.2216399120,arega2023automatic,kalkhof2023m3d,qiu2023qcresunet,Qiu_2025_MIA}. Semi-detached methods work separately but are specifically tailored for a particular family of segmentation approaches\vincenzo{~\cite{Aresta_MIDL,Jebril_2025_IEEE_Access,lin2022novel}}. In contrast, independent QC methods are fully detached from any segmentation model, which makes them versatile and applicable across various segmentation frameworks\vincenzo{~\cite{audelan2019unsupervised,fournel2021medical,galati2021efficient,Li_2022_MIA, liu2019alarm,Jin_2024_MedPhys,robinson2018realtime,specktor2025segqc,valindria2017reverse,wang2020deep}}. We focus on independent QC approaches due to their flexibility and adaptability, as they can be used without being tied to a specific model.

Among detached QC, metric-specific approaches typically focus on either classifying segmentation masks using qualitative scores (e.g., good/bad)~\cite{kohlberger2012evaluating}, or regressing quantitative scores, such as the Dice Score \vincenzo{~\cite{fournel2021medical,Li_2022_MIA,liu2019alarm,qiu2023qcresunet,robinson2018realtime,Qiu_2025_MIA}}. However, these methods are limited by the difficulty of gathering a sufficiently representative set of annotations covering the full spectrum of varying segmentation qualities~\cite{qiu2023qcresunet,robinson2018realtime}, and their inability to handle unbounded metrics (e.g., the Hausdorff Distance)~\cite{fournel2021medical}.
\vincenzo{Aiming to address the reliance on large annotated datasets, a subset of metric-specific approaches~\cite{liu2019alarm,Jin_2024_MedPhys} exploits the observation that high-quality segmentations inherently share common shape properties, and that these properties can be captured within a learned latent space. To this end, they employ a variational autoencoder (VAE) to learn the manifold of high-quality segmentation shapes, and a downstream regressor then operates on the learned embeddings to directly predict a quality score. While this design leverages the representational power of generative reconstruction to encode segmentation shape priors, it remains inherently metric-specific: the regressor is trained to predict a specific metric. Therefore, extending it to another measure requires retraining of the regression head from scratch.}

Reconstruction-based QC techniques~\cite{valindria2017reverse,galati2021efficient,wang2020deep} are detached methods that circumvent the limitations of metric-specific approaches. These techniques generate a pseudo-ground-truth (pGT) mask associated with a given image and its corresponding predicted segmentation, enabling the estimation of quality scores for the predicted segmentation.
Early reconstruction-based QC approaches~\cite{valindria2017reverse,Robinson2019Automated}, relied on atlas propagation strategies.
These registration-based methods assess segmentation quality by measuring the spatial overlap between the predicted mask and a set of reference atlas images. The underlying assumption is that a high-quality prediction will align well with at least one of the atlas images. However, the strategy depends on accurate image registration, which can be computationally expensive~\cite{galati2021efficient} and is prone to failure. Moreover, it requires access to annotated ground truth data at inference time.

More recent reconstruction-based-QC approaches also assume that high-quality segmentations lie in a common space. However, instead of assuming spatial alignment~\cite{valindria2017reverse,Robinson2019Automated} (i.e., Euclidean space), \vincenzo{similarly to earlier metric-specific techniques~\cite{liu2019alarm}, these methods~\cite{galati2021efficient,wang2020deep} build on the assumption} that high-quality ground truth masks lie within a learnable latent manifold. While more efficient and robust, these methods suffer from two critical limitations. First, because they rely on distance-based retrieval to find the closest sample point to the predicted segmentation in the learned space, issues can arise when the segmentation to be controlled is very poor and is far from the underlying normative distribution of high-quality segmentations. In such cases, this distance-based matching may fail, resulting in pseudo-ground truths (pGTs) that no longer resemble the actual ground truth, ultimately leading to unreliable quality estimates.  The latter problem may be exacerbated by the fact that state-of-the-art learning-based QC techniques operate in 2D~\cite{galati2021efficient,wang2020deep,Li_2022_MIA,Jin_2024_MedPhys,fournel2021medical}. Previous studies~\cite{fournel2021medical} have shown that performing QC at the slice level yields better results and provides finer granularity. However, the loss of three-dimensional information, which carries relevant geometric properties of a segmentation mask, can be detrimental to the sampling process. For example, segmented 2D masks of the heart's left ventricle should appear larger in the basal slices compared to the axial slices.

In this work, we leverage the advantages of 2D learning-based reconstruction-based QC techniques while addressing their limitations. We propose a novel sampling strategy that learns to generate high-quality pGTs from a diffusion-based generative model, guided by a rich embedding of visual and spatial cues. 
By combining 3D contextual information with visual cues extracted from the input image, the proposed distance-based retrieval with conditional sampling is better suited to recover pGTs from poor segmentations while preserving consistency with the patient-specific anatomy represented in the image.

%By injecting 3D contextual information, the proposed distance-based retrieval with conditional sampling is better suited to recover pGTs from poor segmentations. \revisiontwo{In parallel, the incorporation of visual context ensures the reconstructed masks remain faithful to the patient anatomy.} This strategy provides a scalable, model-agnostic QC solution. 

Table \ref{tab:structural_comparison} summarizes the main characteristics of our proposed approach and compares them to state-of-the-art QC methods that build upon the principle that high-quality ground truth masks lie within a learnable latent manifold. \revisiontwo{In the comparison, we consider six features: 1) the nature of the generation process (Stochastic); the encoding of 2) Image Context and 3) spatial location (3D  Context); 4) the ability of the model to incorporate additional input information to guide or constrain the reconstruction process (Conditioning); 5) the model's Self-Configuring nature; and 6) whether it is Metric-Agnostic.}% Here, Conditioning refers to the ability of the model to incorporate additional input information to guide or constrain the reconstruction process.}
%summarizes the main characteristics of our proposed approach and compares them to state-of-the-art QC methods that build upon the principle that high-quality ground truth masks lie within a learnable latent manifold.}% applicable across organs, datasets, and imaging modalities.

\begin{table*}[t]
\centering
\caption{Structural Comparison between nnQC and recent manifold-based QC methods. %\revisiontwo{'Conditioning' refers to the model's ability to guide the generative process using external spatial and/or visual information from the input data.}
}
\label{tab:structural_comparison}
%\resizebox{\columnwidth}{!}{%
\begin{tabular}{l|c|c|c|c|c|c}
\hline
\vincenzo{\textbf{Model}} & \vincenzo{\textbf{Stochastic}} & \vincenzo{\textbf{Image Context}} & \vincenzo{\textbf{3D Context}} & \vincenzo{\textbf{Conditioning}} & \vincenzo{\textbf{Self-Configuring}} & \vincenzo{\textbf{Metric-Agnostic}} \\
\hline
\vincenzo{Liu et al. \cite{liu2019alarm}} & \vincenzo{\textcolor{green}{\checkmark}} & \vincenzo{\textbf{\textcolor{red}{\texttimes}}} & \vincenzo{\textcolor{green}{\checkmark}} & \vincenzo{\textbf{\textcolor{red}{\texttimes}}}  & \vincenzo{\textcolor{red}{\texttimes}} & \vincenzo{\textbf{\textcolor{red}{\texttimes}}} \\
\hline
\vincenzo{Jin et al. \cite{Jin_2024_MedPhys}} & \vincenzo{\textcolor{green}{\checkmark}} & \vincenzo{\textbf{\textcolor{red}{\texttimes}}} & \vincenzo{\textcolor{green}{\checkmark}} &   \vincenzo{\textbf{\textcolor{red}{\texttimes}}} & \vincenzo{\textcolor{red}{\texttimes}} & \vincenzo{\textbf{\textcolor{red}{\texttimes}}} \\
\hline
\vincenzo{Galati et al. \cite{galati2021efficient}} & \vincenzo{\textbf{\textcolor{red}{\texttimes}}} & \vincenzo{\textbf{\textcolor{red}{\texttimes}}} & \vincenzo{\textbf{\textcolor{red}{\texttimes}}} & \vincenzo{\textbf{\textcolor{red}{\texttimes}}} & \vincenzo{\textbf{\textcolor{red}{\texttimes}}} & \vincenzo{\textcolor{green}{\checkmark}} \\
\hline
\vincenzo{Wang et al. \cite{wang2020deep}} & \vincenzo{\textcolor{green}{\checkmark}} & \vincenzo{\textcolor{green}{\checkmark}} & \vincenzo{\textbf{\textcolor{red}{\texttimes}}} & \vincenzo{\textbf{\textcolor{red}{\texttimes}}} &  \vincenzo{\textcolor{green}{\checkmark}}& \vincenzo{\textcolor{green}{\checkmark}} \\
\hline
\vincenzo{nnQC (Ours)} & \vincenzo{\textcolor{green}{\checkmark}} & \vincenzo{\textcolor{green}{\checkmark}} & \vincenzo{\textcolor{green}{\checkmark}} & \vincenzo{\textcolor{green}{\checkmark}} & \vincenzo{\textcolor{green}{\checkmark}} & \vincenzo{\textcolor{green}{\checkmark}} \\
\hline
\end{tabular}
\end{table*}

\subsection{Image Synthesis}
%\revision{Discuss: VAEs, Diffusion Models}
Image synthesis, powered by generative modeling, is a powerful tool in medical imaging that is used in numerous applications~\cite{pinaya2022brain,tudosiu2024realistic,bercea2023mask,fernandez2024generating,gupta2024topodiffusionnet}. While earlier approaches primarily relied on Variational Autoencoders (VAEs)~\cite{rezende2015variational} and Generative Adversarial Networks (GANs)~\cite{goodfellow2014generative}, recent trends favor diffusion models (DMs) due to their superior training stability (better than GANs) and high-fidelity sample quality~\cite{ho2020denoising,song2020denoising,rombach2021highresolution} (better than VAEs).
However, the high computational demands of DMs, operating in the image space, limit their scalability in medical imaging applications. Latent Diffusion Models (LDMs)~\cite{rombach2021highresolution} overcome this by performing the diffusion process in a learned latent space, typically using a spatially-aware VAE or VAE-GAN. This approach enables LDMs to sample more effectively than traditional VAEs while preserving essential structural information in a compact space. This is particularly useful for reconstruction-based QC, where severely corrupted masks may deviate from plausible segmentations. LDMs sampling process can guide the output towards realistic, high-quality reconstructions, avoiding the risk of producing overly smoothed or implausible results~\cite{rombach2021highresolution,valindria2017reverse}.

Only a few previous works have explored the usage of DMs for segmentation mask generation. Fernández et al.~\cite{fernandez2024generating} use a VAE-GAN–based mask generator to condition an LDM for image synthesis. Gupta et al.~\cite{gupta2024topodiffusionnet} propose a DM to generate topologically accurate masks for subsequent image generation. In both scenarios, the generated masks are an intermediate step towards the final goal of image synthesis.

In this work, we build on the LDM framework for image synthesis proposed by~\cite{fernandez2024generating}, and we extend it and adapt it to address a slightly different setup. In our case, we aim at generating segmentation masks (i.e. pseudo ground truths) guided by an input segmentation mask, whose quality is to be assessed, and the original input image.

\subsection{Generalist and Specialist Frameworks}
%\revision{Discuss methods like MedSAM, etc with a focus on nnQC. We should thing of a better title for the section.
Recent advances in medical image segmentation have led to the emergence of generalist models capable of segmenting a wide range of anatomical structures from different protocols and imaging modalities with minimal manual intervention (e.g., prompts, scribbles, or bounding boxes)~\cite{butoi2023universeg,MedSAM,wong2024scribbleprompt}. 

\begin{figure*}[!t]
    \centering
    \includegraphics[width=\linewidth]{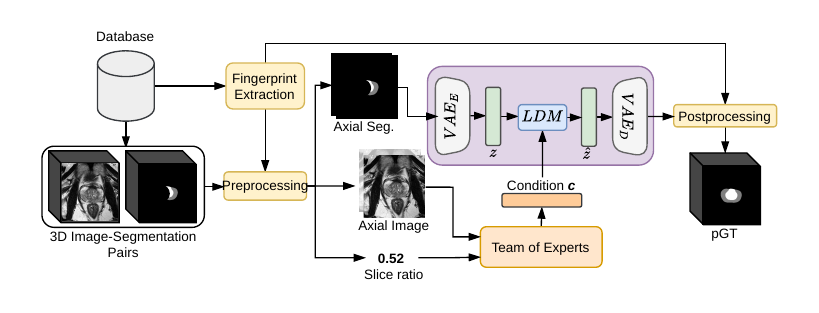}
    \caption{
      The nnQC framework. For a 3D image–segmentation pair, dataset-specific \textit{fingerprints} are extracted and used to preprocess it. Each axial segmentation slice and its corresponding 2D image are passed to the Team of Experts (ToE), which produces conditioning embeddings $c$ for the latent diffusion process. A VAE-GAN maps the 2D segmentation to be quality checked into a latent space of high-quality segmentations from which a DDIM-based Latent Diffusion Model (LDM) generates a pseudo-ground truth ($pGT$). A postprocessing restores the $pGT$ to its original space.
    }
    \label{fig:nnqc_pipeline}
\end{figure*}

Alongside them, specialist models, most notably nnU-Net~\cite{isensee2021nnu}, remain highly competitive, often surpassing generalist models. nnU-Net exemplifies a “one-for-all” model paradigm, where its architecture is self-configuring and retrained from scratch for each new dataset. Despite requiring full retraining, its strong performance, automation of preprocessing and hyperparameter tuning, and ease of use have made it a de facto standard in the field \cite{isensee2024nnu}. Both generalist and specialist approaches have enabled rapid, large-scale deployment of medical image segmentation.

In this work, we take inspiration from nnU-Net's self-adaptation strategy, integrating dataset-specific fingerprints and, thus, removing the need for manual tuning. In this way, we address the bottleneck that QC represents at the moment to medical image segmentation pipeliness, by offering a robust, scalable solution to QC that can be easily adapted across organs, datasets, and imaging techniques.

\section{Method}
%\subsection{Problem Formulation}
Given an image $I \in \mathbb{R}^{H\times W}$, with $H$ its height and $W$ its width, and its associated segmentation \textit{S} generated by an arbitrary segmentation model, we aim to perform segmentation QC by generating a pseudo-ground truth segmentation, $pGT_I^S$ that approximates the real but unknown ground truth segmentation, $GT_I$. The pGT then serves as a reference for computing the quality score of \textit{S}  using an arbitrary quality metric $M(S, pGT^S_I)$, such that $M(S, pGT^S_I) \simeq M(S, GT_I)$. 

We address the QC problem by learning to sample from a learned manifold of GT segmentations (Sec.~\ref{subsec:vae}). To generate $pGT_I^S$, we rely on a latent diffusion process that is formulated as a restoration task, where a latent diffusion model (LDM) is trained to denoise corrupted masks under the guidance of $S$ (Sec.~\ref{subsec:ldm}). Central to nnQC, the learning process is conditioned by a set of embeddings, referred to as \textit{opinions}, which are generated by a conditioning mechanism, denoted the Team of Experts (ToE) module (Sec.~\ref{subsec:experts}). The ToE introduces 3D spatial awareness, represented by the relative spatial position of the axial slice (referred to as the \textit{slice ratio}), and anatomical information derived from visual features extracted from the image $I$. 

Within the training and inference of the proposed framework (Sec.~\ref{sec:traintest}), we integrate the usage of fingerprint adaptation to ensure adaptability across organs, datasets, and imaging modalities (Sec.~\ref{subsec:fingerprints}) Figure~\ref{fig:nnqc_pipeline} presents an overview of the proposed nnQC framework.

\subsection{Manifold of Good Quality Segmentations}
\label{subsec:vae}
nnQC builds on the hypothesis that good-quality segmentations lie on a common manifold~\cite{galati2021efficient,liu2019alarm,wang2020deep}. We learn such a high-quality manifold from a Variational Autoencoder (VAE) trained in an adversarial fashion, i.e., a VAE-GAN~\cite{fernandez2024generating,pinaya2022brain,rombach2021highresolution}, using ground truth (GT) masks. 

The 2D spatial VAE-GAN acts as a shallow autoencoder, applying a non-aggressive downsampling to the input GT mask dimension by a factor of 3: %The spatial VAE learns to compress high-quality, one-hot-encoded binary GT masks into a latent space \( Z \in \mathbb{R}^{C \times H/3 \times W/3} \). 
The encoder $VAE_E$ learns to compress the high-dimensional input mask $x \in \mathbb{R}^{H \times W \times 1}$ into a low-dimensional latent representation $z = \mathcal{E}(x)$ where $z \in \mathbb{R}^{\frac{H}{8} \times \frac{W}{8} \times C}$. This dimensionality follows standard LDM compression configurations~\cite{pinaya2022brain,rombach2021highresolution}, ensuring a compact latent space for efficient sampling while preserving sufficient high-frequency spatial details for accurate mask reconstruction. We set the latent space channel $C=2$ to accommodate the expression of high-level segmentation features, while preserving a good trade-off of spatial relevance in the compressed latent space. As in~\cite{fernandez2024generating}, the spatial VAE is optimized with the following loss:
\begin{align}
\mathcal{L}_{\text{VAE}} = \;
&\lambda_{\text{KLD}}\mathcal{L}_{\text{KLD}}\bigl(\text{VAE}_E(S) \parallel \mathcal{N}(0,1)\bigr) \\ 
&+ \lambda_{perc}\mathcal{L}_{perc}\bigl(S, \hat{S}\bigr) \nonumber\\
&+ \lambda_{adv}\mathcal{L}_{adv}\bigl(D(S), D(\hat{S})\bigr) \nonumber\\
&+ \lambda_{Dice}\mathcal{L}_{Dice}\bigl(S, \hat{S}\bigr) \nonumber
\end{align}
where \( S \) is an input segmentation mask (i.e., a GT mask), \( \hat{S} \) is the reconstructed segmentation, and $\text{VAE}_E$ the encoder of the VAE. \( \mathcal{L}_{\text{KLD}} \) is the Kullback-Leibler divergence loss that forces the latent space $\text{VAE}_E(S)$ to be normally distributed,  \( \mathcal{L}_{perc} \) denotes a perceptual loss \cite{yang2018low}, \( \mathcal{L}_{Dice} \) represents the generalized Dice Loss, and \( \mathcal{L}_{adv} \) is a patch-GAN adversarial loss~\cite{gur2020hierarchical} obtained by forwarding synthetic and real segmentations through a patch-GAN discriminator \( D \) \cite{fernandez2024generating}. We choose \( \mathcal{L}_{Dice} \) as it allows the VAE to learn the spatial relationships among different classes in the input segmentation \cite{galati2021efficient}; \( \mathcal{L}_{perc} \) and \( \mathcal{L}_{adv} \) are also included due to their proven effectiveness in improving reconstruction quality \cite{fernandez2024generating,rombach2021highresolution}. The different $\lambda$ coefficients serve as weights modulating the contribution of individual loss to $\mathcal{L}_{\text{VAE}}$.

\subsection{Latent Diffusion Models for Pseudo Ground Truth Generation}
\label{subsec:ldm}
Once the manifold of high-quality segmentations $Z$ is learned, current approaches generate the pseudo-ground truths $pGT_I^S$ by decoding $Z$ in a deterministic fashion~\cite{galati2021efficient} or through iterative search of the learned latent space~\cite{wang2020deep}. In nnQC, $pGT_I^S$ is generated through a latent diffusion process that operates in the compressed, normative latent space $Z$ learned by the VAE-GAN (Sec.~\ref{subsec:vae}).

We cast the diffusion process as a \textit{restoration} task~\cite{bercea2023mask}: a latent diffusion model (LDM)~\cite{rombach2021highresolution} is trained to iteratively denoise corrupted masks under the guidance of an auxiliary signal, namely the segmentation mask $S$ to be quality controlled. For this purpose, the initial latent representation \( z_0 \in Z\) is corrupted by injecting an \textit{imperfect} segmentation. 

To simulate a wide range of realistic \textit{imperfect} segmentations, we synthetically corrupt the available GT masks using random morphological perturbations (see Sec.~\ref{sec:exps}).

Following the objective function in~\cite{ho2020denoising}, for a given timestep $t \in [0, T]$ of the reverse diffusion process, the LDM is trained to minimize
\begin{equation}\label{eq:ldm}
    \mathcal{L}_{\text{LDM}} = \left\| \epsilon - \epsilon_{\theta}(z_{t,S}; c) \right\|^2_2, 
\end{equation}
where $\epsilon_{\theta}$ is the learned function to predict the true noise $\epsilon \sim \mathcal{N}(0,1)$ from $z_{t,S}$, the latent representation $z_t$ corrupted with an imperfect segmentation $S$, given a condition $c$. In nnQC, we design the condition $c$ to encode 3D spatial information and visual anatomical features derived from $I$ to guide the denoising process. The mechanism to build $c$, which we refer to as \textit{Team of Experts}, is presented in the following.

\begin{figure*}[!t]
    \centering
    \includegraphics[width=\textwidth]{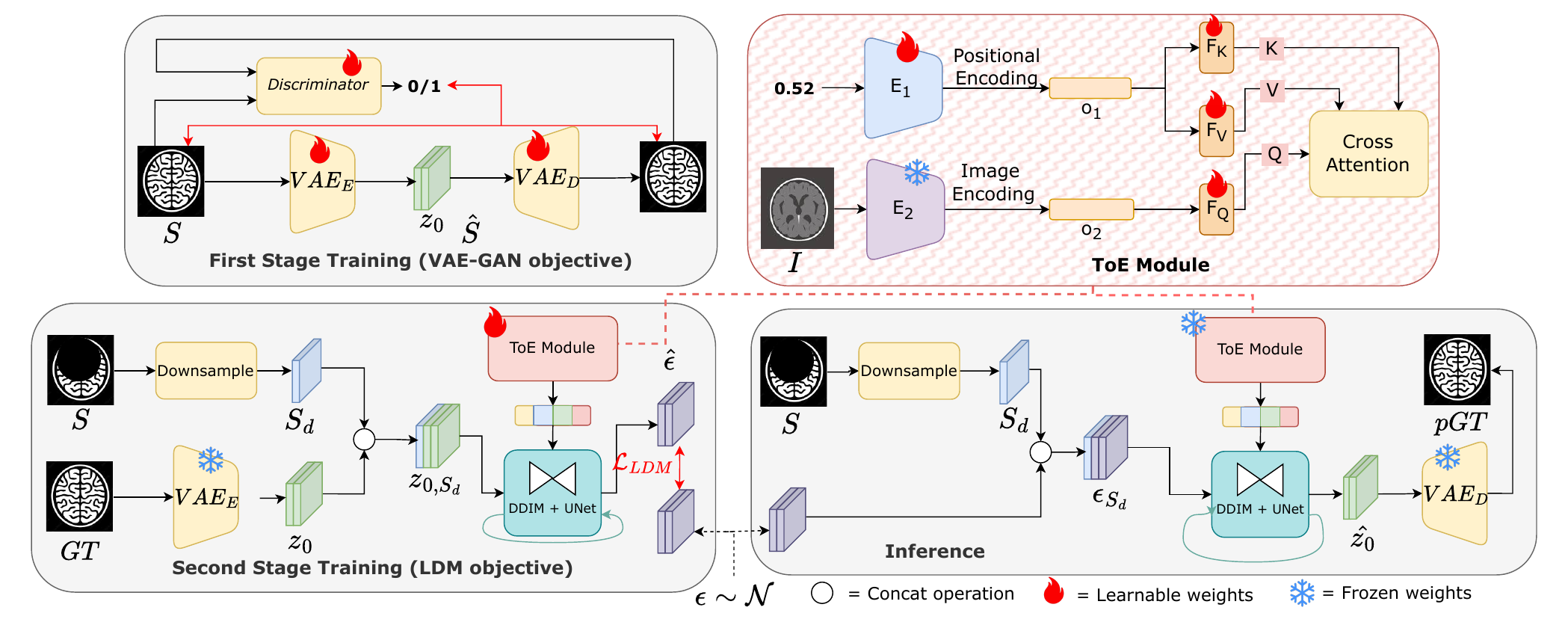}
    \caption{Two-stage training and inference workflow. At the first stage (top left), the VAE is trained adversarially (i.e., a VAE-GAN) to learn a rich latent space of high-quality segmentations (i.e., GTs). During the training's second stage (bottom left), the LDM learns to reconstruct noise conditioned by embeddings from the \textit{Team of Experts} (ToE) module (top right). The ToE's positional embedding is jointly optimized with the LDM. At inference (bottom right), Gaussian noise and $S_d$ are fed into the LDM; the ToE-generated condition $c$ guides the LDM to recover \(z_0\), which is decoded by VAE$_D$ to generate the pGT.} 
    \label{fig:ldm_stage}
\end{figure*}

\subsection{Team of Experts: Dual Embeddings for LDM Conditioning}
\label{subsec:experts}
We enforce nnQC to sample from the learned high-quality segmentations manifold by conditioning the LDM on two complementary feature sets extracted from $I$, guiding the sampling process for generating pGTs. Each feature set, or \textit{opinion}, is derived from a specialized feature extractor, or \textit{expert}. We denote the set of features as the \textit{Team of Experts} (ToE) (Figure~\ref{fig:ldm_stage}).

nnQC deals with 2D image–segmentation pairs extracted from 3D volume pairs, which have been proven to yield better results~\cite{fournel2021medical}, but lack three-dimensional information that may be important to the sampling process.
We address this limitation by injecting 3D contextual information into the conditioning. 

To this end, we introduce Expert \( E_1 \) to encode the relative spatial position of the axial slices, expressed as a \textit{slice-to-volume ratio} in the range \([0, 1]\), into a fixed-dimensional embedding vector, \( o_1 \). This is achieved through a lightweight Multi-Layer Perceptron (MLP) that receives the slice-to-volume ratio as input and outputs the positional embedding, \( o_1 \). \( E_1 \) is jointly trained with the LDM (Eq.~\ref{eq:ldm}) rather than separately. In this way, we ensure that the learned positional embedding space remains semantically aligned with the latent manifold learned by the VAE-GAN. Moreover, the shared optimization scheme prevents the collapse of the latent space and promotes spatial conditioning throughout the generation process.

Previous works~\cite{wang2020deep} have shown that utilizing information from $I$ yields better reconstruction than relying solely on the information conveyed by $S$~\cite{galati2021efficient}. nnQC follows a similar approach. However, rather than integrating information from $I$ by reconstructing the $(I, S)$ pair~\cite{wang2020deep}, which can add complexity to the training process, nnQC encodes visual features from $I$ within a vector $o_2$. This is achieved through the definition of Expert \( E_2 \), which leverages a pretrained CLIP-like vision encoder to extract high-level semantic features from $I$. 

By employing UniMedCLIP \cite{khattak2024unimed}, a vision encoder pretrained on a large set of medical and clinical data, we expect the resulting embedding, $o_2$, to capture anatomical information from $I$. \revisiontwo{This relies on the assumption that UniMedCLIP provides semantically meaningful and sufficiently stable representations even when image quality degradations are severe enough to affect segmentation accuracy. Since $E_2$ is used only as a pretrained feature extractor, its vision encoder weights are frozen during the joint optimization of $E_1$ and the LDM.}

To dynamically balance the opinions from the ToE, we utilize a Cross-Attention mechanism \cite{chen2021crossvit,rebain2022attention} that serves as a \textit{dynamic switch}. It assigns appropriate importance to each opinion, and it generates a unified conditioning vector \( c \). This is achieved by projecting both \( o_1 \) and \( o_2 \) using linear layers \( F_Q, F_K, F_V \) to produce a query \( Q = F_Q(o_1) \), key \( K = F_K(o_2) \), and value \( V = F_V(o_2) \). Afterwards, the conditioning vector $c$ (Eq.~\ref{eq:ldm}) is thus obtained as the Cross-Attention vector  
\begin{equation}
    c = \text{Attention}(Q, K, V) = \text{softmax} \left( \frac{Q K^T}{\sqrt{d_k}} \right) V, 
\end{equation}
where \( d_k \) is the dimensionality of the keys, and fed as a condition to the diffusion process. 

% s 

\subsection{Two-stage Training and Inference Workflows}
\label{sec:traintest}

The nnQC framework follows a two-stage training design. Figure~\ref{fig:ldm_stage} illustrates the two-stage training and inference workflows. 

\subsubsection{Training}  In the first stage, we train the VAE-GAN using adversarial training \cite{fernandez2024generating,rombach2021highresolution}. The frozen VAE's encoder, i.e., VAE$_E$, is then used as part of the LDM's training, during the second stage. 

We adopt a Denoising Diffusion Implicit Model (DDIM) \cite{song2020denoising}, which constructs a non-Markovian forward process that preserves the same training objective as Denoising Diffusion Probabilistic Models (DDPMs) \cite{ho2020denoising}, but allows for a more efficient deterministic sampling procedure. For the LDM's internal model, we use a conditional UNet architecture \cite{rombach2021highresolution,ronneberger2015u} as the network that learns the denoising process. During the second stage of training, we set the number of diffusion steps to $T=1000$. We set $\epsilon \in \mathbb{R}^{2 \times H/3 \times W/3}$ to be consistent with the dimensionality of $Z$, as defined in Section~\ref{subsec:vae}. Similarly, synthetically generated \textit{imperfect} segmentations $S$ are rescaled to the [0,1] range and downsampled to \( S_d \in \mathbb{R}^{1 \times H/3 \times W/3} \). The resulting $S_d$ is concatenated with the sampled $z_0$, forming the input \( z_{0,S_d} \in \mathbb{R}^{3 \times H/3 \times W/3} \) for the diffusion UNet. \vincenzo{
Crucially, the optimization of the LDM is shared with the trainable components of the Team of Experts (ToE). Specifically, the gradients derived from minimizing $\mathcal{L}_{LDM}$ (Eq. 2) are backpropagated to update both the weights of the denoising U-Net and the parameters of the positional expert $E_1$, along with the linear projection layers of the cross-attention mechanism ($F_Q, F_K, F_V$). This joint training strategy ensures that the generated conditioning embeddings $c$ are semantically aligned with the latent diffusion space, preventing latent collapse and enabling effective guidance for the restoration task. 
}

\subsubsection{Inference} We leverage the efficiency of DDIM sampling, which offers a flexible trade-off between sample quality and generation speed, substantially reducing computational costs~\cite{song2020denoising}. While DDIMs typically use 50 steps \cite{song2020denoising}, we empirically reduce the number of sampling steps to $T=20$ as two factors mitigate the generation complexity: 1) the concatenation of the input mask with the input noise, which injects a strong bias into the process, and 2) limiting the image domain to binary masks with pixels constrained to $\{0,1\}$.

During the sampling process, inference mirrors training: a randomly sampled Gaussian noise \(\epsilon \in \mathbb{R}^{2 \times H/3 \times W/3} \) is concatenated with the rescaled and downsampled segmentation mask $S$ to be quality controlled, yielding the input noise \(\epsilon_{S_d} \in \mathbb{R}^{3 \times H/3 \times W/3}\). This input is denoised by the trained UNet to reconstruct the latent representation \( z_0 \) by reversing the diffusion process. Finally, the denoised latent sample is decoded by the VAE decoder, \( \text{VAE}_D \), to produce \( pGT^S_I \).  %The full procedure is summarized in Algorithm \ref{alg:nnqc}, while 

\subsection{Fingerprints for Self-Adaptable QC}
\label{subsec:fingerprints}
Inspired by nnUNet~\cite{isensee2021nnu}, we use \revisiontwo{dataset} fingerprints to enable our framework to self-adapt to various data types and conditions. We define the \textit{fingerprints} as a set of key characteristics that describe the input dataset: the median voxel spacing of subject volumes, the median size of foreground regions, image orientation, intensity ranges specific to each modality, and the number of unique segmentation classes. These fingerprints form the basis for dataset-specific adaptations during both data pre-processing and post-processing, as well as at the network/model level. 
At pre-processing, the fingerprints guide image rescaling. The median voxel spacing and the median cropped volume size are used to standardize the image dimensions (\(256 \times 256\)), while image contrast is scaled based on the 0.5 and 99.5 percentile intensity values within the foreground regions~\cite{isensee2021nnu}, ensuring modality-specific normalization. The rescaled images are aligned to a predefined orientation (right, anterior, superior, \textit{``RAS"}).
During post-processing, the fingerprints serve to restore the original resolution.
% %, yielding uniform samples for training. This preprocessing guarantees normalized input images for \( E_2 \)'s encoding and standardizes the resolution of the (\textit{I, S}) pairs to \(256 \times 256\). Moreover, it enforces a consistent foreground area across slices and harmonizes subjects to a \vincenzo{homogeneous volumetric space based on the previously estimated median voxel spacing.} 

At the network level, the \revisiontwo{dataset} fingerprints allow the selection of the number of input and output channels of the VAE's first and last layers using the number of segmentation labels in the dataset. Unlike the intensive fingerprint-based adaptation process in nnUNet \cite{isensee2021nnu}, we leverage the intrinsic adaptability of LDMs to operate within a predefined image space~\cite{rombach2021highresolution}. As a result, we use a homogeneous latent size across all datasets, which simplifies training while preserving flexibility.

\section{Experiments and Results}
\label{sec:exps}
\subsection{Experimental design and setup}

\subsubsection{Datasets} 
We conduct experiments \revisiontwo{across} \vincenzo{15} datasets covering seven organ types, three imaging modalities - magnetic resonance imaging (MRI), computed tomography (CT), and ultrasound (US) - and a varying number of annotated structures (labels). We use six datasets from the Medical Segmentation Decathlon (MSD) challenge \cite{antonelli2022medical}, encompassing Spleen (61 CT scans, 1 label), Prostate (48 MRI volumes, 2 labels), Heart (30 Late Gadolinium Enhancement MRI volumes, 1 label), Liver (210 CT volumes, 1 label), Pancreas (420 CT volumes, 1 label), and Hippocampus (394 MRI volumes, 2 classes); the ACDC (150 Short Axis Cardiac MRI volumes, 3 labels), M\&M-2 (360 Short Axis Cardiac MRI volumes, 3 labels), and CAMUS datasets (500 US scans, 3 labels) for heart segmentation; KiTS 2021 \cite{KITS2021} (300 CT scans, 1 label) for kidney segmentation; CHAOS 2021 \cite{CHAOS2021} for abdominal organ segmentation from MR images (40 MRI volumes, 1 label per organ: kidney, spleen, and liver); PROSTATE-X \cite{armato2018prostatex} (346 MRI volumes, 2 labels) for prostate segmentation; \vincenzo{ and AbdomenCT-1K~\cite{ma2021abdomenct} (Abd1K-CT; 1000 CT volumes, 1 label for spleen and for liver).}

% \vincenzo{
% %The number of classes reported above always refers to foreground regions only, with the background excluded. 
% .}

%For the cross-dataset experiment, we include the prostate PROSTATE-X dataset (MRI) and the cardiac M\&M-2 dataset (MRI).

\subsubsection{Benchmarks}
We consider three reconstruction-based QC baselines for comparison: 
\textit{(1) Galati et al.}~\cite{galati2021efficient} a deterministic reconstructor based on a Convolutional Autoencoder, which reconstructs segmentation masks to restore their original shape; \textit{(2) Liu et al.}~\cite{liu2019alarm} a two-stage regressor that uses a VAE trained to learn the normative good-quality manifold of GTs and an MLP that processes the features generated by the latent space obtained from the reconstructed segmentation to predict the a pseudo Dice score; and \textit{(3) Wang et al.}~\cite{wang2020deep}, a VAE that processes the channel-wise concatenation of image-segmentation pairs, and adjusts their compressed embeddings in the latent-space using a stochastic iterative search. \vincenzo{ Unlike other baselines which show missing training setups and architectural choices,  which fundamentally limited our ability to directly reproduce and benchmark against their method in our experimental setup, our benchmark selection is motivated by the public availability of implementation details.}

\subsubsection{Evaluation Metrics} We assess performances using the Pearson correlation (\textit{r}) and the Mean Absolute Error (MAE) between the predicted pseudo-quality scores (using a pGT) and real quality scores (using the available GT). We use the Dice-Sørensen Coefficient (DSC) and the 95\% Hausdorff Distance (HD95) as quality metrics. In the cross-dataset experiment, we use the MAE between predicted and real scores. 

\subsubsection{Setup \& Implementation Details}
We adopt an 80--20\% training--testing split at the subject level. We use GT labels to learn the manifold of GT segmentations (Sec.~\ref{subsec:vae}) at the first stage of training. During the second stage, we simulate segmentations of varying quality by corrupting the GT from the considered datasets through synthetic degradations \vincenzo{\cite{Aresta_MIDL,Jebril_2025_IEEE_Access}}, allowing the LDM to learn how to recover good-quality segmentation masks from degraded ones. \vincenzo{This choice is motivated by the need to generate a diverse and unbiased training distribution of mask quality levels. While one could in principle use the outputs of a specific segmentation model to obtain masks at varying quality, this would introduce an architectural bias, as the framework would be exposed only to the failure modes and artefact patterns characteristic of that particular model, ultimately compromising its ability to generalise across different segmentation sources~\cite{robinson2018realtime,fournel2021medical}. By contrast, corrupting ground-truth masks directly through synthetic degradations produces a broad and model-agnostic spectrum of quality levels, an approach already adopted in the medical imaging literature~\cite{Aresta_MIDL,Jebril_2025_IEEE_Access}.} The GTs are degraded to five distinct levels, corresponding to uniformly spread DSC intervals of [0.05-0.10), [0.10-0.25), [0.25-0.50), [0.50-0.75), and [0.75-0.95].
\vincenzo{We implement the degradation pipeline including: (i) \textit{Morphological Perturbations} with random erosion and dilation operations (kernel sizes 3-7) simulate systemic under- and over-segmentation errors, mirroring the boundary uncertainty often observed in nnU-Net and MedSAM \cite{MedSAM} when tissue contrast is low; (ii) \textit{Cutout and Random Holes} by randomly mask out regions within the organ to simulate false negatives and mimic the 'missed region' artifacts; (iii) \textit{Additive Noise (False Positives)} with the injection of random mask blobs into the background to simulate false positives, a frequent failure mode in Transformer-based models (e.g., SwinUNETR \cite{hatamizadeh2021swin}).}
During testing, GTs are also subject to degradation through the same procedure. The resulting test set comprises a total of 9,370 2D slices. We aggregate the 2D predictions on 3D volumes to compute the evaluation metrics at the subject level.

For the benchmark models, we follow the guidelines of the respective studies. For \cite{galati2021efficient} and \cite{wang2020deep}, we rely on the publicly available codebase provided by the authors, while for \cite{liu2019alarm}, we implemented their pipeline and model architecture as described in their study. For both nnQC and benchmark models, we train one model per organ, i.e., for CHAOS we do not train a single model across all abdominal organs, but rather a separate model for each organ. 

All code is developed in Python 3.10, using the MONAI library and PyTorch 2.0. Training experiments are run on an 80 GB NVIDIA A100, \vincenzo{brought by the French national cluster IDRIS on Jean-Zay machines} with a 12.4 CUDA version (average GPU memory consumption with a batch size of 32 is around 32Gb).

\begin{figure*}
    \centering
    \includegraphics[width=\linewidth]{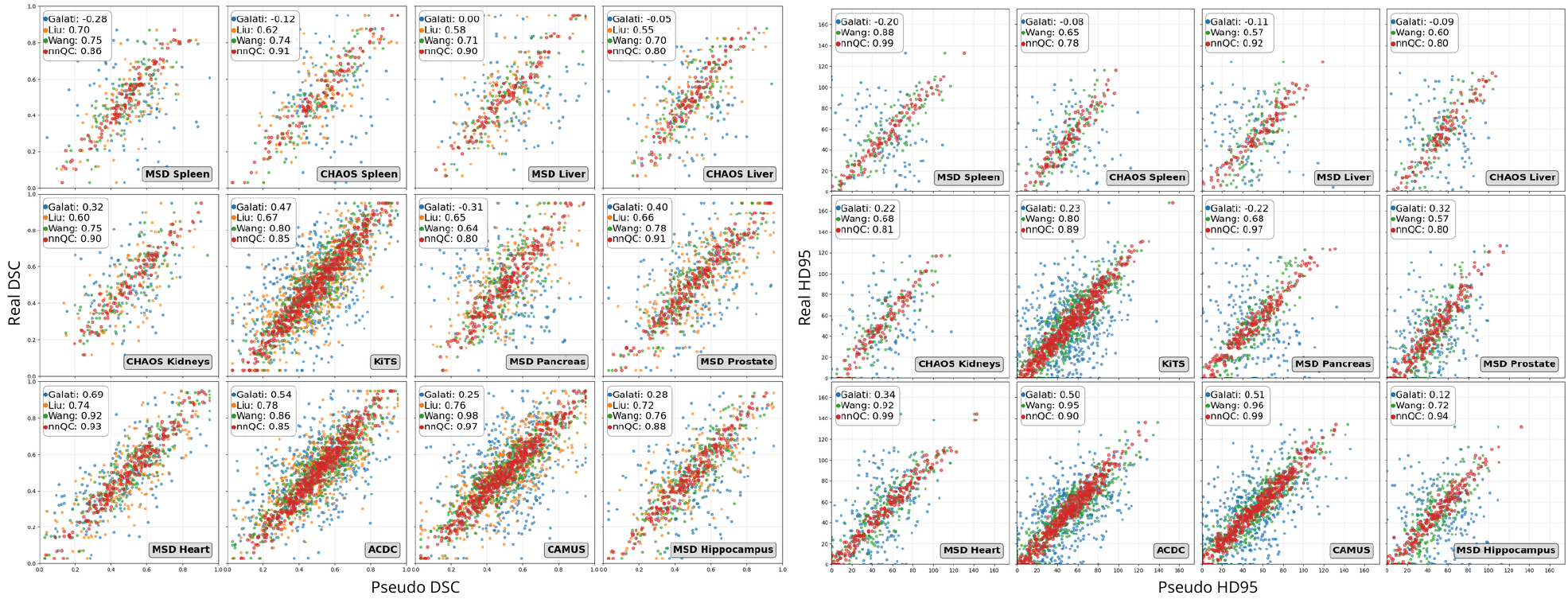}
    \caption{Pearson correlation (\textit{r})  between the predicted pseudo-quality scores and real scores (DSC and HD95) across different organs, modalities, and datasets. HD95 is not estimated for Liu et al~\cite{liu2019alarm} as their model is designed to predict pseudo DSCs.}
    \label{tab:correlation}
\end{figure*}
\begin{figure*}
    \centering
    \includegraphics[width=\linewidth]{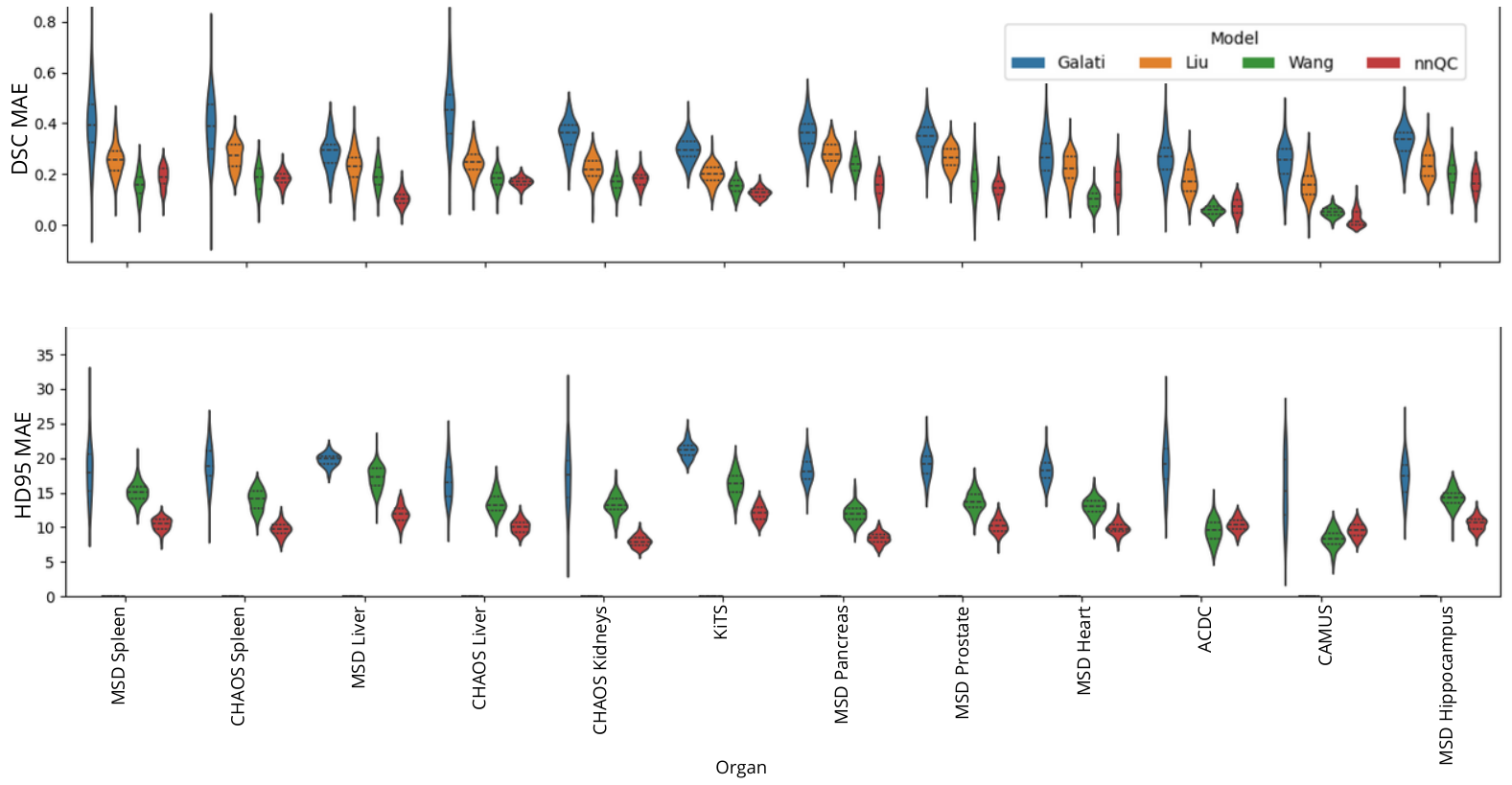}
    \caption{Mean Absolute Error (MAE) distribution across different organs, modalities, and datasets. The MAE is measured as the difference between the predicted pseudo-quality scores and real scores (DSC and HD95). As in Fig~\ref{tab:correlation}, HD95 is not estimated for Liu et al~\cite{liu2019alarm}.}
    \label{tab:maes}
\end{figure*}

\subsection{Results}

\subsubsection{Benchmark Study}  \label{sec:benchmark}
\revisiontwo{We assess the performance of nnQC and compare it against benchmark methods on synthetically degraded masks from 12 datasets. \revisiontwo{PROSTATEx, M\&M-2, and Abdomen1K-CT are excluded from this study and held out for cross-dataset generalization assessment.}}  Figure~\ref{tab:correlation} reports the Pearson correlation coefficient ($r$) and Figure~\ref{tab:maes} the MAE. 

Models relying solely on mask information, such as~\cite{galati2021efficient}, perform poorly with an average DSC MAE of \(0.36 \pm 0.10\), HD95 MAE of \(19.2 \pm 3.3\), and low correlations (DSC $r=0.21 \pm 0.23$, HD95 $r=0.13 \pm 0.12$). Liu et al.~\cite{liu2019alarm} achieves a better, but yet limited performance with an average DSC MAE of \(0.25 \pm 0.05\) and moderate correlations (mean DSC $r=0.62 \pm 0.11$). Both approaches exhibit broad, heavy-tailed error distributions, reflecting limited robustness across organs. Instead, models that also use information from the original image report a competitive performance, as observed in the results from
Wang et al.~\cite{wang2020deep} (mean DSC $r=0.77 \pm 0.09$, mean HD95 $r=0.78 \pm 0.12$, average DSC MAE of 0.16$\pm0.15$ and average HD95 MAE of 11.2$\pm3.84$), confirming the importance of also encoding information from the original image. %, 

Nonetheless, nnQC consistently reports a better performance across organs, both in terms of $r$ (mean DSC \(r = 0.89 \pm 0.03\) and HD95 \(r = 0.94 \pm 0.02\)) and MAE (DSC MAE \(0.12 \pm 0.06\) and HD95 MAE \(9.54 \pm 2.33\)) outperforming the baseline models. For instance, Wang et al. achieve the best correlations in heart-related datasets, such as ACDC and CAMUS; however, their performance degrades in other datasets, including MSD Pancreas and MSD Liver. This can be explained by the fact that Wang's original model~\cite{wang2020deep} has been conceived for heart segmentation QC. Instead, nnQC has been designed to be easily adapted across organs, datasets, and imaging techniques, which is reflected in its consistent performance across different scenarios. 

\subsubsection{Cross-dataset generalization}  
We assess nnQC's generalization capabilities through a cross-dataset evaluation using an out-of-distribution (OOD) test set, i.e., one different from the training sets. Models trained on MSD Prostate and ACDC are evaluated on PROSTATEx and M\&M-2 heart, with 380 and 160 subjects, respectively, totaling 9226 axial slices. Additionally, to further assess generalization across different organs and imaging modalities, we evaluate models trained on MSD Spleen and MSD Liver on the corresponding OOD splits from the Abd1K-CT. Table \ref{tab:cross_dataset} presents the obtained results in terms of the MAE between predicted and real DSC \vincenzo{and HD95}, and compares them against Wang et al.~\cite{wang2020deep}, the best competing model in the benchmark study.

\begin{table}[b]
\centering
\caption{\vincenzo{Cross-dataset performance. Models are trained on MSD Prostate, ACDC, MSD Spleen and MSD Liver. Bold denotes best. }}
\label{tab:cross_dataset}
\resizebox{\columnwidth}{!}{%
\begin{tabular}{l l cc}
\hline
& & \textbf{DSC MAE} & \vincenzo{\textbf{HD95 MAE (mm)}} \\
\hline
\multirow{2}{*}{PROSTATEx } & Wang & $0.20 \pm 0.06$ & \vincenzo{$9.84 \pm 4.12$} \\
& nnQC & $\mathbf{0.09 \pm 0.03}$ & \vincenzo{$\mathbf{5.23 \pm 2.67}$} \\
\hline
\multirow{2}{*}{M\&M-2} & Wang & $0.15 \pm 0.08$ & \vincenzo{$7.62 \pm 3.88$} \\
& nnQC & $\mathbf{0.10 \pm 0.03}$ & \vincenzo{$\mathbf{4.41 \pm 2.03}$} \\
\hline
\multirow{2}{*}{\vincenzo{Abd1K-CT Spleen}} & \vincenzo{Wang} & \vincenzo{$0.17 \pm 0.09$} & \vincenzo{$12.43 \pm 5.21$} \\
& \vincenzo{nnQC} & \vincenzo{$\mathbf{0.08 \pm 0.04}$} & \vincenzo{$\mathbf{6.87 \pm 3.14}$} \\
\hline
\multirow{2}{*}{\vincenzo{Abd1K-CT Liver}} & \vincenzo{Wang} & \vincenzo{$0.14 \pm 0.14$} & \vincenzo{$18.62 \pm 7.83$} \\
& \vincenzo{nnQC} & \vincenzo{$\mathbf{0.13 \pm 0.08}$} & \vincenzo{$\mathbf{9.24 \pm 4.56}$} \\
\hline
\end{tabular}}
\end{table}

nnQC reports low MAE values that closely match those achieved in the in-distribution (ID) dataset (see Section~\ref{sec:benchmark}), where a MAE of 0.14$\pm$0.04 was recorded for the MSD Prostate dataset, and 0.07$\pm$0.04 for the ACDC dataset. Notably, we found that the MAE in the PROSTATEx dataset is lower than that of the ID data, underscoring nnQC's ability to generalize to unseen OOD data. In contrast, Wang et al.~\cite{wang2020deep} show a drop in performance when exposed to OOD data, as evidenced by an increase in MAE compared to the values obtained from the ID data (0.16$\pm$0.07 for MSD Prostate and 0.06$\pm$0.02 for ACDC). The generalization trend is further confirmed on the abdominal datasets: nnQC achieves a DSC MAE of $0.08 \pm 0.04$ and $0.13 \pm 0.08$ on Abd1K-CT Spleen and Liver, respectively, consistently outperforming Wang et al. across both organs and both metrics. \revisiontwo{Statistical significance was assessed using paired t-tests, confirming that nnQC significantly outperformed Wang et al.~\cite{wang2020deep} across all evaluated OOD datasets and metrics ($p<0.05$).}

The superior generalization capabilities of nnQC can be attributed to its use of UniMedCLIP  representations, obtained from a vision encoder pre-trained on large and diverse medical datasets. This pre-training makes nnQC more robust to domain shifts. Furthermore, the use of relative positional encodings helps disambiguate spatial structures, ensuring consistent performance even in the presence of OOD data.

\subsubsection{Model ranking}  

We assess whether the pseudo-quality scores produced through nnQC can be used for model ranking. To that end, we consider three state-of-the-art medical image segmentation frameworks, nnUNet~\cite{isensee2021nnu}, MedSAM~\cite{MedSAM}, and SwinUNETR~\cite{hatamizadeh2021swin}, along with two reference baselines emulating a perfect model and a low-performance one. For the first one, we use the GT masks. For the second one, we rely on an atlas-based segmentator using ANTs~\cite{avants2009advanced} with five image-segmentation pairs from the training set as atlases, employing a joint-fusion policy to segment the unseen images. We generate segmentations across three cardiac datasets (MSD Heart, ACDC, and CAMUS) encompassing different imaging techniques and semantic labels. We use the pseudo-DSC obtained from nnQC \revisiontwo{and Wang et al.~\cite{wang2020deep}} to rank the five models and compare these rankings with those obtained using the GT. Ranking agreement is measured with Kendall's \(\tau\) test (Table~\ref{tab:rank}).

\revisiontwo{Overall, nnQC achieves the highest average Kendall's $\tau$ over the three datasets ($\tau=0.87$ vs. $\tau=0.80$ for Wang et al.), indicating better agreement with the GT-based ranking.} In MSD Heart (late gadolinium enhancement MRI), nnQC perfectly reproduces the ranking that would be obtained using the ground truth. For ACDC and CAMUS (MRI and US), the rankings are reproduced with the exception of two swaps between MedSAM and SwinUNETR (ACDC) and between nnUNet and SwinUNETR (CAMUS), corresponding to a Kendall’s \(\tau=0.80\). These discrepancies are likely due to subtle differences in performance between the models. To validate this hypothesis, we performed \revisiontwo{t-tests} on the real DSC distributions for each pair of models involved in a rank swap to assess whether there is a significant difference between \revisiontwo{their DSC performance}. The t-test yielded p-values of 0.704 (MedSAM vs. SwinUNETR in ACDC) and 0.112 (nnUNet vs. SwinUNETR in CAMUS), indicating that the observed rank swaps occur in settings where the performance differences are not statistically significant.

\revisiontwo{Wang et al. achieves perfect ranking agreement on CAMUS, which is consistent with its original design for myocardium segmentation, a structure characterized by relatively low shape variability and stable ring-like anatomy. Similar properties are observed in CAMUS, where cardiac structures exhibit limited inter-subject geometric variation. However, on MSD Heart and ACDC, where the segmentation targets present more complex anatomical configurations and higher variability, Wang et al. shows reduced agreement with the GT-based ranking. These observations are consistent with the findings of the benchmark study and further support the ability of nnQC to provide more reliable ranking across datasets and imaging modalities.}

\begin{table}[t]
\centering 
\caption{\revisiontwo{Rankings obtained from nnQC and Wang et al. are compared against the GT-based ranking. Kendall's $\tau$ measures agreement with the GT ranking.}}
\resizebox{\columnwidth}{!}{%
%\begin{tabular}{l|l|c|c|c}
\begin{tabular}{l|l*{3}{>{\centering\arraybackslash}m{1.6cm}}}
\hline\hline
\textbf{Dataset} & \textbf{Model} & \textbf{GT} & \revisiontwo{\textbf{Wang et al.}} & \textbf{nnQC} \\
\hline
                 & GT        & 1 & \revisiontwo{1} & 1 \\
                 & nnUNet    & 2 & \revisiontwo{3} & 2 \\
 MSD Heart       & MedSAM    & 3 & \revisiontwo{2} & 3 \\
                 & SwinUNETR & 4 & \revisiontwo{4} & 4 \\
                 & ANTs      & 5 & \revisiontwo{5} & 5 \\
                 \cline{2-5}
                 & \cellcolor{gray!15}\textbf{Kendall's $\tau$} & \cellcolor{gray!15}\textbf{1.00} & \cellcolor{gray!15}\revisiontwo{\textbf{0.80}} & \cellcolor{gray!15}\textbf{1.00} \\
\hline
                 & GT        & 1 & \revisiontwo{1} & 1 \\
                 & nnUNet    & 2 & \revisiontwo{4} & 2 \\
 ACDC            & MedSAM    & 3 & \revisiontwo{2} & 4 \\
                 & SwinUNETR & 4 & \revisiontwo{3} & 3 \\
                 & ANTs      & 5 & \revisiontwo{5} & 5 \\
                 \cline{2-5}
                & \cellcolor{gray!15}\textbf{Kendall's $\tau$} & \cellcolor{gray!15}\textbf{1.00} & \cellcolor{gray!15}\revisiontwo{\textbf{0.60}} & \cellcolor{gray!15}\textbf{0.80} \\
\hline
                 & GT        & 1 & \revisiontwo{1} & 1 \\
                 & nnUNet    & 2 & \revisiontwo{2} & 3 \\
 CAMUS           & SwinUNETR & 3 & \revisiontwo{3} & 2 \\
                 & MedSAM    & 4 & \revisiontwo{4} & 4 \\
                 & ANTs      & 5 & \revisiontwo{5} & 5 \\
                 \cline{2-5}
                 & \cellcolor{gray!15}\textbf{Kendall's $\tau$} & \cellcolor{gray!15}\textbf{1.00} & \cellcolor{gray!15}\revisiontwo{\textbf{1.00}} & \cellcolor{gray!15}\textbf{0.80} \\
\hline
\multicolumn{2}{r}{\textbf{Average Kendall’s \(\tau\)}} & \textbf{1.00} & \revisiontwo{\textbf{0.80}} & \textbf{0.87} \\
\hline\hline
\end{tabular}
}
\label{tab:rank}
\end{table}

\subsubsection{Ablation study}
We conduct an ablation study to understand the role of the \textit{opinions} from the ToE module in the framework's performance. In particular, we study performance as we remove the cross-attention module and disable one expert at a time to condition the LDM. For the study, we consider two datasets: CAMUS, where nnQC performs best, and CHAOS Liver, where nnQC performance is the lowest (Figures~\ref{tab:correlation} and~\ref{tab:maes}). Table~\ref{tab:ablation} reports the obtained results.

The ablation studies reveal consistent results across datasets, highlighting the importance of 3D spatial information from positional encodings for model performance. Using positional encodings alone yields the highest DSC $r$, while image encodings have a lower performance on their own. This is likely due to subtle changes in appearance (i.e., texture and intensity), making image encodings less informative. Nonetheless, the full model performs best, indicating that the information from both \textit{experts} is complementary and enhances nnQC's performance.

\subsubsection{Analysis of the visual expert}
\revisiontwo{Given the central role of the visual expert $E_2$ in conditioning the reconstruction process, we investigate the impact of the underlying vision encoder and the robustness of UniMedCLIP under simulated MRI corruptions.}

\revisiontwo{We first evaluate the impact of the visual encoder on pGT reconstruction quality by replacing UniMedCLIP with two alternatives, the general-domain CLIP~\cite{radford2021learning} and the earlier medical-specific MedCLIP \cite{wang2022medclip}, on the CAMUS (ultrasound) and CHAOS Liver (CT) datasets. Table \ref{tab:encoder_ablation} reports DSC \textit{r} and DSC MAE. }

\revisiontwo{CLIP shows the lowest performance across both datasets, likely reflecting the domain gap between natural and medical images. MedCLIP improves over CLIP, supporting the benefit of medical-domain pretraining. UniMedCLIP achieves the highest DSC correlation and lowest DSC MAE on both CAMUS and CHAOS Liver, indicating that its domain-aligned representations provide more effective conditioning for pGT reconstruction. These findings are consistent with the conclusions of the original UniMedCLIP work~\cite{khattak2024unimed}, suggesting that large-scale medical pretraining improves the robustness and anatomical relevance of the extracted features.}

%Having established the importance of the visual expert, we further investigated its specific architectural design. We compared UniMedCLIP with the general-domain CLIP  and the earlier medical-specific MedCLIP \cite{wang2022medclip}, evaluating the reconstruction quality using these three encoders on the CAMUS (ultrasound) and CHAOS Liver (CT/MRI) datasets. The results are summarized in Table \ref{tab:encoder_ablation}. As observed in these results, the standard CLIP model struggles significantly, likely because its training domain is dominated by natural images. Conversely, while MedCLIP offers a substantial improvement by adapting to the medical domain, UniMedCLIP consistently yields the highest correlation and lowest error rates across both modalities. Ultimately, these findings align with the conclusions of the original UniMedCLIP study, confirming that its robust, domain-aligned semantic features are crucial for guiding the diffusion model to an accurate and anatomically consistent reconstruction.}

\begin{table}[t]
\centering
\caption{Ablation study on the CHAOS Liver and CAMUS datasets. Bold denotes best performance.}
\label{tab:ablation}
\resizebox{\columnwidth}{!}{%
\begin{tabular}{l l ccc}
\hline
\textbf{Dataset} & \textbf{ToE Configuration} & \textbf{DSC r} & \textbf{HD95 r} & \textbf{DSC MAE} \\
\hline
\vincenzo{CHAOS Liver} & \vincenzo{no conditioning} & \vincenzo{0.66} & \vincenzo{0.59} & \vincenzo{0.32} $\pm$ \vincenzo{0.18} \\
CHAOS Liver & with Image Encoding & 0.72 & 0.75 & 0.18 $\pm$ 0.08 \\
CHAOS Liver & with Positional Encoding      & \textbf{0.85} & 0.75 & 0.20 $\pm$ 0.04 \\
CHAOS Liver & Full Model   & 0.80 & \textbf{0.80} & \textbf{0.17 $\pm$ 0.03} \\
\hline
\vincenzo{CAMUS} & \vincenzo{no conditioning} & \vincenzo{0.71} & \vincenzo{0.43} & \vincenzo{0.27} $\pm$ \vincenzo{0.25} \\
CAMUS  & with Image Encoding & 0.86 & 0.88 & 0.12 $\pm$ 0.07 \\
CAMUS  & with Positional Encoding & \textbf{0.90} & 0.92 & 0.11 $\pm$ 0.04 \\
CAMUS  & Full Model   & 0.89 & \textbf{0.97} & \textbf{0.05 $\pm$ 0.04} \\
\hline
\end{tabular}
}
\end{table}

\begin{table}[t]
\centering
\caption{\revisiontwo{Ablation of the vision expert ($E_2$) architecture.}}
\label{tab:encoder_ablation}
\resizebox{\columnwidth}{!}{%
\begin{tabular}{llcc}
\toprule
\revisiontwo{\textbf{Dataset}} & \revisiontwo{\textbf{Visual Encoder}} & \revisiontwo{\textbf{DSC $r$ $\uparrow$}} & \revisiontwo{\textbf{DSC MAE $\downarrow$}} \\ 
\midrule
\multirow{3}{*}{\revisiontwo{\textbf{CAMUS}}} 
& \revisiontwo{CLIP \cite{radford2021learning}}       & \revisiontwo{0.68} & \revisiontwo{$0.21 \pm 0.10$} \\ 
& \revisiontwo{MedCLIP \cite{wang2022medclip}}    & \revisiontwo{0.83} & \revisiontwo{$0.14 \pm 0.06$} \\ 
& \revisiontwo{\textbf{UniMedCLIP (Ours)}} & \revisiontwo{\textbf{0.89}} & \revisiontwo{\textbf{0.05 $\pm$ 0.03}} \\ 
\midrule
\multirow{3}{*}{\revisiontwo{\textbf{CHAOS Liver}}} 
& \revisiontwo{CLIP \cite{radford2021learning}}       & \revisiontwo{0.55} & \revisiontwo{$0.27 \pm 0.14$} \\ 
& \revisiontwo{MedCLIP \cite{wang2022medclip}}    & \revisiontwo{0.76} & \revisiontwo{$0.19 \pm 0.08$} \\ 
& \revisiontwo{\textbf{UniMedCLIP (Ours)}} & \revisiontwo{\textbf{0.80}} & \revisiontwo{\textbf{0.17 $\pm$ 0.05}} \\ 
\bottomrule
\end{tabular}%
}
\end{table}

\begin{figure}[t]
    \centering
    \includegraphics[width=\linewidth]{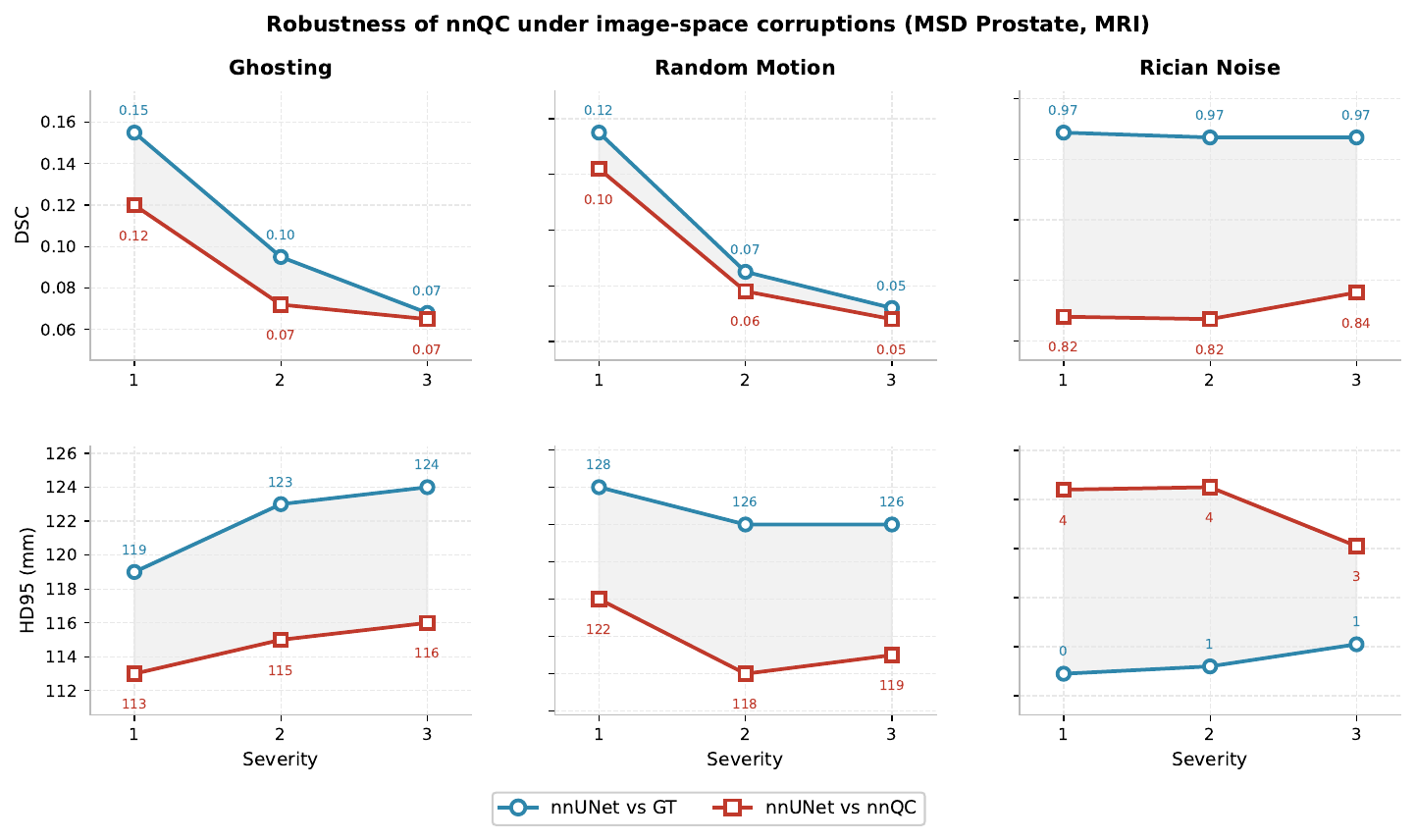}
    \caption{\revisiontwo{Robustness analysis of the UniMedCLIP-based visual expert $E_2$ within nnQC on nnU-Net segmentations obtained from ROOD-MRI corrupted MSD Prostate images.}}
    \label{fig:rood}
\end{figure}

%\subsubsection{Robustness to image-space corruptions}
%\label{subsec:rood}

\revisiontwo{
Beyond encoder choice, we also assessed whether the visual expert could provide semantically meaningful and sufficiently stable representations under input image degradation. To this end, we conducted a robustness experiment using ROOD-MRI~\cite{boone2023rood}, a framework for simulating MRI artifacts and generating out-of-distribution (OOD) samples. We considered three corruption types representative of common MRI degradations, namely Ghosting, Random Motion, and Rician Noise, each applied at three increasing severity levels (1 to 3). Experiments were performed on the MSD Prostate dataset, using a trained nnU-Net as the segmentation model under evaluation.}

\revisiontwo{As expected, nnU-Net degraded substantially under severe corruptions: under Ghosting and Random Motion artefacts, the DSC dropped to approximately 0.07 at severity 3, while HD95 reached around 125 mm. Under Rician Noise, nnU-Net remained relatively stable, with DSC close to 0.97, suggesting that this corruption does not strongly alter the underlying anatomical information. When using nnQC as a surrogate quality measure, by comparing the generated pGT against the nnU-Net predictions, we observed consistently high agreement under Rician Noise conditions, with DSC above 0.80 and HD95 below 5 mm across all severity levels. These results suggest that the combination of the visual expert $E_2$ and the positional expert $E_1$ provides sufficiently stable guidance for the reconstruction process when the underlying anatomical structure remains preserved in the conditioning image. The results are shown in Figure~\ref{fig:rood}.
}

\begin{figure*}[t] 
    \centering 
    \includegraphics[width=0.95\linewidth]{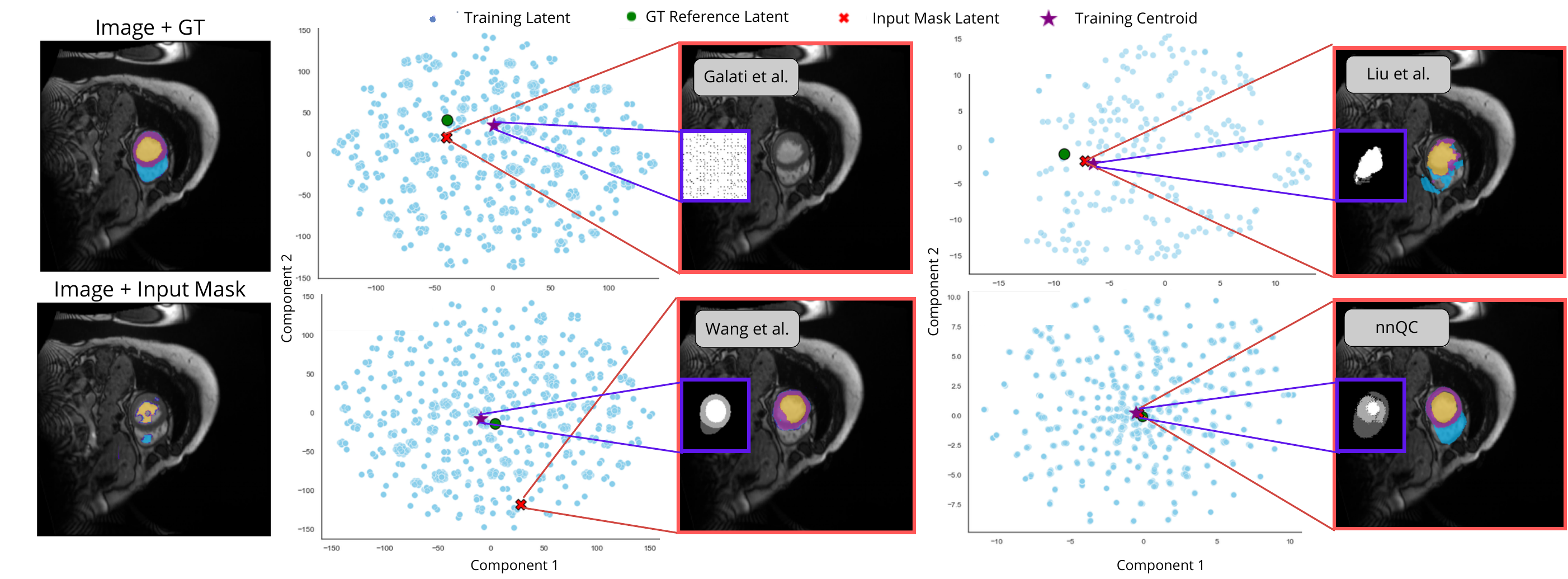} 
    \caption{
    Learned normative manifolds and generated pGTs from a low-quality input segmentation from the ACDC dataset. The first column shows the GT and a low-quality segmentation overlaid in the original image. The following blocks display the latent spaces learned by different QC methods, the reconstructed pGTs (red box), and the reconstructed centroids (purple box). The projected manifolds are obtained using t-SNE~\cite{maaten2008visualizing}.}
    \label{fig:qualit}
\end{figure*}

\subsubsection{Qualitative latent-retrieval analysis}
\label{sec:qualit}
\vincenzo{Lastly, we study the learned latent representations across nnQC and the baselines. Figure~\ref{fig:qualit} shows 2D projections of the different normative learned manifolds and their respective centroids in ACDC. Using a randomly selected sample for QC, we visualize its location and that one of the reference GT in the latent space, as well as the corresponding reconstructed pGT. Additionally, we display the reconstructed centroid, as it provides insights into the model’s implicit \textit{idea} of the represented domain~\cite{Higgins2016betaVAELB}, or, in this case, its average understanding of anatomical variability.}

\vincenzo{In Galati et al.~\cite{galati2021efficient} and Liu et al.~\cite{liu2019alarm} the centroids exhibit abnormal reconstructions, which may reflect on the quality of the learned latent representation.
%are then reflected in the models' anatomically implausible pGTs. 
Specifically, in \cite{galati2021efficient}, the reconstucted consists of a flat mask dominated by one class with scattered artifacts from other classes, lacking any relevant semantic information. As a result, when faced with a poor-quality segmentation, the model collapses into a blank pGT. Similarly, in \cite{liu2019alarm}, the reconstructed centroid mask displays fragmented and inconsistent contours, leading to an incomplete and erroneous pGT in the example. Instead,
%when the model samples from a low-quality input. 
Wang et al.~\cite{wang2020deep} present a centroid that corresponds to a segmentation mask with a well-defined anatomical shape, indicating that the latent space encodes a strong anatomical prior, which in turn enables the model to generate anatomically plausible shapes. 
Nonetheless, this smooth ``average'' shape suggests a learned latent representation that cannot fully capture the high variability across shapes, which may stem from the limited size of the latent encoding (i.e., $\mathbb{R}^{16}$). The plausible but \textit{anatomically incorrect} pGT in Figure~\ref{fig:qualit} (where the right ventricle class is not generated) can be further explained by the iterative sampling mechanism implemented in~\cite{wang2020deep}, which stops once it retrieves a plausible shape. This behavior suggests that the model's conditioning on the intensity image is insufficient to guide the sampling process effectively.}

\vincenzo{In contrast, nnQC's reconstructed centroid can be described as a topological template of the considered anatomy. Although the contours are noisier, the reconstructed centroid preserves the spatial relationship between anatomical structures (e.g., left ventricle enclosed by myocardium and myocardium adjacent to right ventricle). Unlike \cite{wang2020deep}, ours captures a more abstract concept of the anatomy, encompassing anatomical variability rather than a concrete shape instance, as a direct consequence of the richer 2D latent space. This latent representation offers a meaningful starting point for the  ToE-conditioned diffusion process, which then refines this anatomical template into subject-specific reconstruction variations, where the sampled pGT closely resembles the corresponding GT (Figure \ref{fig:qualit}).}

\subsubsection{Computational cost analysis}
\vincenzo{We compare the training and inference time of nnQC against the baseline methods. Training time is reported per epoch (with consistent batch sampling across datasets), while inference time is measured by processing all slices of 10 randomly selected subjects per dataset and averaging the results (Table \ref{tab:computational_cost}).}

\vincenzo{On an NVIDIA A100 GPU, the total training time per epoch for nnQC is 204.5s, which is higher than Wang et al. (92.0s) and Liu et al. (148.1s). This increased training cost reflects the added complexity of our Latent Diffusion Model (LDM) architecture and two-stage training pipeline (where the training time refers to the total of both stages). This reflects a trade-off with the improved performance and robustness observed in our results across diverse organs. For inference, while nnQC has higher latency (404ms per sample) than Wang et al. (32ms) and Galati et al. (12ms) due to iterative sampling, it remains well within practical subsecond latency for offline quality control workflows. This reflects the trade-off between computational cost and reconstruction quality.}

\begin{table}[t]
\centering
\caption{Computational cost comparison among the proposed benchmarks and nnQC.}
\label{tab:computational_cost}
\resizebox{\columnwidth}{!}{%
\begin{tabular}{l|c|c|c|c}
\hline
\textbf{Model} & \textbf{Liu et al.} & \textbf{Galati et al.} & \textbf{Wang et al.} & \textbf{nnQC (Ours)} \\
\hline
Training time per epoch (s) & 148.1 & 23.6 & 92.0 & \vincenzo{204.5} \\
\hline
Inference time per sample (s) & 0.226 ± 0.167 & 0.012 ± 0.009 & 0.032 ± 0.017 & 0.404 ± 0.221 \\
\hline
\end{tabular}
}
\end{table}

\section{Conclusion}
\vincenzo{In this work, we introduced \textit{nnQC}, a model- and metric-agnostic quality control framework for segmentation masks that generates reliable pseudo-ground truths through a novel sampling strategy. At its core, nnQC features a \textit{Team of Experts (ToE)} module that independently processes the input image and relative axial position by using %treating 
cross-attention as a dynamic mechanism to balance their contributions %\vincenzo{The proposed design of nnQC handles both correct and incorrect input segmentation masks, spanning across diverse segmentation qualities.} 
Furthermore, nnQC extracts dataset-specific \textit{fingerprints} that allow for automatic adaptation to a wide range of anatomical structures and imaging modalities. Extensive experiments %Different evaluations 
across twelve datasets, seven organs and three image modalities demonstrated that nnQC outperforms state-of-the-art methods, confirming itself as a versatile QC solution, that can robustly handle high- and low-quality segmentations across organs and imaging modalities.}

\vincenzo{We have, however, identified some pending limitations. First, external experiments indicate that nnQC struggles with complex multi-organ segmentations, where recovering accurate inter-class topological relationships becomes difficult. For instance, Figure~\ref{fig:chaos_fails} illustrates a pGT failure on a multi-organ scenario (CHAOS dataset). We hypothesize this behavior arises from the large spatial separation among organ classes, which hinders nnQC from forming a coherent, normative segmentation \textit{template}. Currently, we circumvent this by using separate models for each organ, but it would be desirable to have a single model to handle QC across all organs in an image.
\begin{figure}[!t]
    \centering
    \includegraphics[width=\linewidth]{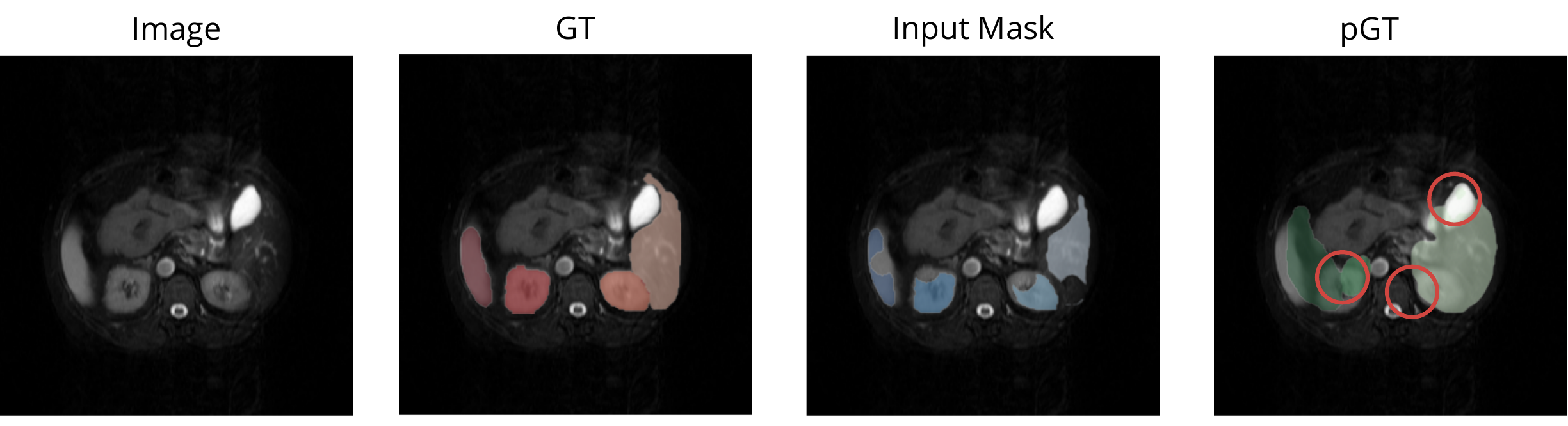}
    \caption{Failed pGT sampling when multiple organ classes are present in the input, in this case, liver, spleen, left and right kidneys. Red circles indicate anatomical inconsistencies in the pGT.}
    \label{fig:chaos_fails}
\end{figure}
Second, the current evaluation excludes highly heterogeneous structures, such as tumors or vascular structures. This choice stems from the inherent difficulty of embedding such structures within a learned normative manifold, as their irregular shapes and heterogeneity prevent including them in a single “good-quality” latent representation. 
Finally, we acknowledge that the relative spatial position encoding employed in nnQC assumes a consistent Field-of-View (FOV) across training and test acquisitions. In scenarios where a significant FOV mismatch exists between datasets, for example, when one acquisition covers the full anatomy, and another captures only a partial slab, the positional encoding may become inconsistent, representing a boundary condition that could affect the reliability of the generated pseudo-ground-truth masks.
To address these limitations, future work may explore: (1) the incorporation of a topological interaction loss to better capture inter-class spatial dependencies; and (2) the extension of nnQC toward a fully 3D formulation, enabling the model to leverage volumetric context for more reliable spatial reasoning.}

%Additionally, we plan to explore the integration of nnQC into real-time segmentation refinement pipelines, expanding its potential for clinical deployment and automated downstream analysis.

%\appendices

%\section*{Appendix and the Use of Supplemental Files}

\section*{Acknowledgment}
This work was supported in part by the French government, through the 3IA Côte d’Azur Investments project (ANR-23-IACL-0001), the ANR-BMBF TRAIN (ANR-22-FAI1-0003-02) and by ANR Fed-Ops (ANR-24-IAS2-0001). Computational resources were provided by GENCI at IDRIS (grant 2025-AD011016543) on the Jean Zay's A100 partition.
Vincenzo Marciano and Maria A. Zuluaga are with EURECOM, Sophia Antipolis, France, and with the School of Biomedical Engineering \& Imaging Sciences, King's College London, UK (email: $\left[\text{marciano,zuluaga}\right]$ @eurecom.fr. )\\
Hava Chaptoukaev is with EURECOM, Sophia Antipolis, France\\
Virginia Fernandez, M. Jorge Cardoso, Michela Antonelli, and Sébastien Ourselin are with the School of Biomedical Engineering \& Imaging Sciences, King's College London, UK \\
Corresponding author: Maria A. Zuluaga

\color{black}
\bibliographystyle{ieeetr}
\bibliography{sample}

@article{boone2023rood,
  title={ROOD-MRI: Benchmarking the robustness of deep learning segmentation models to out-of-distribution and corrupted data in MRI},
  author={Boone, Lyndon and Biparva, Mahdi and Forooshani, Parisa Mojiri and Ramirez, Joel and Masellis, Mario and Bartha, Robert and Symons, Sean and Strother, Stephen and Black, Sandra E and Heyn, Chris and others},
  journal={NeuroImage},
  volume={278},
  pages={120289},
  year={2023},
  publisher={Elsevier}
}

@inproceedings{song2020denoising,
  title={{Denoising diffusion implicit models}},
  author={Song, Jiaming and Meng, Chenlin and Ermon, Stefano},
  booktitle={{International Conference on Learning Representations (ICLR)}},
  year={{2021}},
  url={{https://openreview.net/forum?id=St1giarCHLP}}
}

@inproceedings{goodfellow2014generative,
  title     = {Generative Adversarial Nets},
  author    = {Goodfellow, Ian and Pouget-Abadie, Jean and Mirza, Mehdi and Xu, Bing and Warde-Farley, David and Ozair, Sherjil and Courville, Aaron and Bengio, Yoshua},
  booktitle = {Advances in Neural Information Processing Systems},
  year      = {2014}
}

@inproceedings{butoi2023universeg,
  title={{Universeg: Universal medical image segmentation}},
  author={Butoi, Victor Ion and Ortiz, Jose Javier Gonzalez and Ma, Tianyu and Sabuncu, Mert R and Guttag, John and Dalca, Adrian V},
  booktitle={Proceedings of the IEEE/CVF International Conference on Computer Vision},
  pages={{6830--6839}},
  year={2023}
}

@article{rezende2015variational,
  title   = {Variational Inference with Normalizing Flows},
  author  = {Rezende, Danilo Jimenez and Mohamed, Shakir},
  journal = {International Conference on Machine Learning},
  year    = {2015},
}

@inproceedings{pinaya2022brain,
  title={{Brain imaging generation with latent diffusion models}},
  author={Pinaya, Walter HL and Tudosiu, Petru-Daniel and Dafflon, Jessica and Da Costa, Pedro F and Fernandez, Virginia and Nachev, Parashkev and Ourselin, Sebastien and Cardoso, M Jorge},
  booktitle={MICCAI workshop on deep generative models},
  year={2022}
}

@article{tudosiu2024realistic,
  title={Realistic morphology-preserving generative modelling of the brain},
  author={Tudosiu, Petru-Daniel and Pinaya, Walter HL and Da Costa, Pedro Ferreira and Dafflon, Jessica and Patel, Ashay and Borges, Pedro and Fernandez, Virginia and Graham, Mark S and Gray, Robert J and Nachev, Parashkev and others},
  journal={{Nature Machine Intelligence}},
  volume={{6}},
  pages={{811--819}},
  year={2024},
  publisher={{Nature Publishing Group}},
  doi={10.1038/s42256-024-00870-7}

}

@conference{gupta2024topodiffusionnet,
  title={TopoDiffusionNet: A Topology-aware Diffusion Model},
  author={Gupta, Saumya and Samaras, Dimitris and Chen, Chao},
  booktitle={{
International Conference on Learning Representations (ICLR)}},
  year={{
2025}},
  url={https://openreview.net/forum?id=ZK1LoTo10R}
}

@inproceedings{liu2019alarm,
  title={{An alarm system for segmentation algorithm based on shape model}},
  author={Liu, Fengze and Xia, Yingda and Yang, Dong and Yuille, Alan L and Xu, Daguang},
  booktitle={Proceedings of the IEEE/CVF International Conference on Computer Vision},
  year={2019}
}

@inproceedings{qiu2023qcresunet,
  title={{QCResUNet: Joint subject-level and voxel-level prediction of segmentation quality}},
  author={Qiu, Peijie and Chakrabarty, Satrajit and Nguyen, Phuc and Ghosh, Soumyendu Sekhar and Sotiras, Aristeidis},
  booktitle={International Conference on Medical Image Computing and Computer-Assisted Intervention},
  year={2023},
}

@inproceedings{hatamizadeh2021swin,
  title={{Swin unetr: Swin transformers for semantic segmentation of brain tumors in mri images}},
  author={Hatamizadeh, Ali and Nath, Vishwesh and Tang, Yucheng and Yang, Dong and Roth, Holger R and Xu, Daguang},
  booktitle={Brain Lesion: Toward Natural Science for Brain Tumor Segmentation (BrainLes 2021)},
  series={Lecture Notes in Computer Science},
  volume={12962},
  pages={272--284},
  year={2021},
}

@inproceedings{bercea2023mask,
  title={Mask, Stitch, and Re-Sample: Enhancing Robustness and Generalizability in Anomaly Detection through Automatic Diffusion Models},
  author={Bercea, Cosmin I and Neumayr, Michael and Rueckert, Daniel and Schnabel, Julia A},
  booktitle={ICML 3rd Workshop on Interpretable Machine Learning in Healthcare (IMLH)},
  year={{2023}},

}

@article{avants2009advanced,
  title={Advanced normalization tools (ANTS)},
  author={Avants, Brian B and Tustison, Nick and Song, Gang and others},
  journal={Insight j},
  volume={2},
  number={365},
  pages={1--35},
  year={2009}
}

@article{khattak2024unimed,
  title={{Unimed-clip: Towards a unified image-text pretraining paradigm for diverse medical imaging modalities}},
  author={Khattak, Muhammad Uzair and Kunhimon, Shahina and Naseer, Muzammal and Khan, Salman and Khan, Fahad Shahbaz},
  journal={arXiv preprint arXiv:2412.10372},
  year={2024}
}

@article{valindria2017reverse,
  title={{Reverse classification accuracy: predicting segmentation performance in the absence of ground truth}},
  author={Valindria, Vanya V and Lavdas, Ioannis and Bai, Wenjia and Kamnitsas, Konstantinos and Aboagye, Eric O and Rockall, Andrea G and Rueckert, Daniel and Glocker, Ben},
  journal={IEEE Transactions on Medical Imaging},
  year={2017},
}

@article{specktor2025segqc,
  title={{SegQC: a segmentation network-based framework for multi-metric segmentation quality control and segmentation error detection in volumetric medical images}},
  author={Specktor-Fadida, Bella and Ben-Sira, Liat and Ben-Bashat, Dafna and Joskowicz, Leo},
  journal={Medical Image Analysis},
  volume={103},
  pages={103638},
  year={2025},
  year={2025},
}

@inproceedings{kalkhof2023m3d,
  title={{M3D-NCA}: Robust 3D Segmentation with Built-in Quality Control},
  author={Kalkhof, John and Mukhopadhyay, Anirban},
  booktitle={{Medical Image Computing and Computer Assisted Intervention -- MICCAI 2023}},
  volume={{14220}},
  pages={{169--178}},
  year={{2023}},
  doi={10.1007/978-3-031-43898-1_17}
}

@article{rombach2021highresolution,
    title={High-Resolution Image Synthesis with Latent Diffusion Models},
  author={{Rombach, Robin and Blattmann, Andreas and Lorenz, Dominik and Esser, Patrick and Ommer, Bj{\"o}rn}},
  booktitle={{Proceedings of the IEEE/CVF Conference on Computer Vision and Pattern Recognition (CVPR)}},
  pages={{10684--10695}},
  year={2022},
  doi={{10.1109/CVPR52688.2022.01042}}
}

@inproceedings{fernandez2024generating,
  title={{Generating multi-pathological and multi-modal images and labels for brain MRI}},
  author={Fernandez, Virginia and Pinaya, Walter Hugo Lopez and Borges, Pedro and Graham, Mark S and Tudosiu, Petru-Daniel and Vercauteren, Tom and Cardoso, M Jorge},
  booktitle={{Deep Generative Models (DGM4MICCAI), MICCAI Workshop}},
  volume={{13609}},
  pages={{117--126}},
  year={2022},
  doi={10.1007/978-3-031-16452-1_12}
}

@article{maaten2008visualizing,
  title={Visualizing data using t-SNE},
  author={Maaten, Laurens van der and Hinton, Geoffrey},
  journal={Journal of machine learning research},
  volume={9},
  number={Nov},
  pages={2579--2605},
  year={2008}
}

@inproceedings{chen2021crossvit,
  title={{CrossViT: Cross-attention multi-scale vision transformer for image classification}},
  author={Chen, Chun-Fu Richard and Fan, Quanfu and Panda, Rameswar},
  booktitle={Proceedings of the IEEE/CVF international conference on computer vision},
  pages={{357--366}},
  year={2021}
}

@article{lin2022novel,
  title={A novel quality control algorithm for medical image segmentation based on fuzzy uncertainty},
  author={Lin, Qiao and Chen, Xin and Chen, Chao and Garibaldi, Jonathan M},
  journal={IEEE Transactions on Fuzzy Systems},
  volume={31},
  number={8},
  pages={2532--2544},
  year={2022},
  publisher={IEEE}
}

@inproceedings{wang2020deep,
  title={{Deep generative model-based quality control for cardiac MRI segmentation}},
  author={Wang, Shuo and Tarroni, Giacomo and Qin, Chen and Mo, Yuanhan and Dai, Chengliang and Chen, Chen and Glocker, Ben and Guo, Yike and Rueckert, Daniel and Bai, Wenjia},
  booktitle={Medical Image Computing and Computer Assisted Intervention--MICCAI 2020: 23rd International Conference, Lima, Peru, October 4--8, 2020, Proceedings, Part IV 23},
  year={2020},
}

@article{ho2020denoising,
  title={{Denoising diffusion probabilistic models}},
  author={Ho, Jonathan and Jain, Ajay and Abbeel, Pieter},
  journal={Advances in neural information processing systems},
  year={2020}
}

@article{gur2020hierarchical,
  title={{Hierarchical patch vae-gan: Generating diverse videos from a single sample}},
  author={Gur, Shir and Benaim, Sagie and Wolf, Lior},
  journal={Advances in Neural Information Processing Systems},
  year={2020}
}

@inproceedings{ronneberger2015u,
  title={{U-net: Convolutional networks for biomedical image segmentation}},
  author={Ronneberger, Olaf and Fischer, Philipp and Brox, Thomas},
  booktitle={Medical image computing and computer-assisted intervention--MICCAI 2015: 18th international conference, Munich, Germany, October 5-9, 2015, proceedings, part III 18},
  year={2015},
}

@article{yang2018low,
  title={{Low-dose CT image denoising using a generative adversarial network with Wasserstein distance and perceptual loss}},
  author={Yang, Qingsong and Yan, Pingkun and Zhang, Yanbo and Yu, Hengyong and Shi, Yongyi and Mou, Xuanqin and Kalra, Mannudeep K and Zhang, Yi and Sun, Ling and Wang, Ge},
  journal={IEEE transactions on medical imaging},
  volume={{37}},
  number={{6}},
  pages={{1348--1357}},
  year={2018},
}

@inproceedings{audelan2019unsupervised,
  title={{Unsupervised quality control of image segmentation based on Bayesian learning}},
  author={Audelan, Beno{\^\i}t and Delingette, Herv{\'e}},
  booktitle={Medical Image Computing and Computer Assisted Intervention--MICCAI 2019: 22nd International Conference, Shenzhen, China, October 13--17, 2019, Proceedings, Part II 22},
  year={2019},
}

@conference{galati2021efficient,
      title={Efficient Model Monitoring for Quality Control in Cardiac Image Segmentation}, 
      author={Francesco Galati and Maria A. Zuluaga},
  booktitle={{Functional Imaging and Modeling of the Heart (FIMH 2021)}},
volume={{12738}},
  pages={{101--111}},
      year={2021}
}

@article{isensee2021nnu,
  title={nnU-Net: Self-adapting Framework for Deep Learning-Based Bioedical Image Segmentation},
  author={Isensee, Fabian and Jaeger, Paul Friedrich and Kohl, Simon AA and Petersen, Jens and Maier-Hein, Klaus H},
  journal={Nature Methods},
  year={2021},
}

@article{antonelli2022medical,
  title={{The medical segmentation decathlon}},
  author={Antonelli, Michela and Reinke, Annika and Bakas, Spyridon and Farahani, Keyvan and Kopp-Schneider, Annette and Landman, Bennett A and Litjens, Geert and Menze, Bjoern and Ronneberger, Olaf and Summers, Ronald M and others},
  journal={Nature communications},
  year={2022},
}

@inproceedings{isensee2024nnu,
  title={nnu-net revisited: A call for rigorous validation in 3d medical image segmentation},
  author={Isensee, Fabian and Wald, Tassilo and Ulrich, Constantin and Baumgartner, Michael and Roy, Saikat and Maier-Hein, Klaus and Jaeger, Paul F},
  booktitle={International Conference on Medical Image Computing and Computer-Assisted Intervention},
  pages={488--498},
  volume={{14999}},
  year={2024},
}

@article{armato2018prostatex,
  title={PROSTATEx Challenges for computerized classification of prostate lesions from multiparametric magnetic resonance images},
  author={Armato III, Samuel G and Huisman, Henkjan and Drukker, Karen and Hadjiiski, Lubomir and Kirby, Justin S and Petrick, Nicholas and Redmond, George and Giger, Maryellen L and Cha, Kenny and Mamonov, Artem and others},
  journal={Journal of Medical Imaging},
  year={2018},
}

@article{bernard2018deep,
  title={Deep Learning Techniques for Automatic MRI Cardiac Multi-structures Segmentation and Diagnosis: Is the Problem Solved?},
  author={Bernard, Olivier and Lalande, Alain and Zotti, Claudio and Cervenansky, Frederic and et al.},
  journal={IEEE Transactions on Medical Imaging},
  year={2018}
}

@inproceedings{wong2024scribbleprompt,
  title={Scribbleprompt: fast and flexible interactive segmentation for any biomedical image},
  author={Wong, Hallee E and Rakic, Marianne and Guttag, John and Dalca, Adrian V},
  booktitle={European Conference on Computer Vision},
  volume={15098},
  pages={3--19},
  year={2024}
}

@inproceedings{Aresta_MIDL,
  author    = {{Aresta, Guilherme and Bogunović, Haris}},
  title     = {{FAZ Segmentation Quality Assessment in OCTA via Denoising Autoencoders and Segmentation Uncertainty Estimation}},
  booktitle = {{Medical Imaging with Deep Learning-Short Papers}},
  year={{2025}}
}

@article{arega2023automatic,
  title={{Automatic uncertainty-based quality controlled T1 mapping and ECV analysis from native and post-contrast cardiac T1 mapping images using Bayesian vision transformer}},
  author={{Arega, Tewodros Weldebirhan and Bricq, St{\'e}phanie and Legrand, Fran{\c{c}}ois and Jacquier, Alexis and Lalande, Alain and Meriaudeau, Fabrice}},
  journal={{Medical image analysis}},
  volume={{86}},
  pages={{102773}},
  year={{2023}}
}

@article{Li_2022_MIA,
  author  = {{Li, Kang and Yu, Lequan and Heng, Pheng-Ann}},
  title   = {{Towards reliable cardiac image segmentation: Assessing image-level and pixel-level segmentation quality via self-reflective references}},
  journal = {{Medical Image Analysis}},
  year    = {{2022}},
  doi     = {10.1016/j.media.2022.102426}
}

@article{Jin_2024_MedPhys,
  author  = {{Jin, Xiyao and Hao, Yao and Hilliard, Jessica and Zhang, Zhehao and Thomas, Maria A and Li, Hua and Jha, Abhinav K and Hugo, Geoffrey D}},
  title   = {{A quality assurance framework for routine monitoring of deep learning cardiac substructure computed tomography segmentation models in radiotherapy}},
  journal = {{Medical physics}},
  volume={{51}},
  number={{4}},
  pages={{2741--2758}},
  year    = {{2024}},
  doi     = {10.1002/mp.16846}
}

@article{ma2021abdomenct,
  title={Abdomenct-1k: Is abdominal organ segmentation a solved problem?},
  author={Ma, Jun and Zhang, Yao and Gu, Song and Zhu, Cheng and Ge, Cheng and Zhang, Yichi and An, Xingle and Wang, Congcong and Wang, Qiyuan and Liu, Xin and others},
  journal={IEEE Transactions on Pattern Analysis and Machine Intelligence},
  volume={44},
  number={10},
  pages={6695--6714},
  year={2021},
  publisher={IEEE}
}

@article{Qiu_2025_MIA,
  author  = {{Qiu, Peijie and Chakrabarty, Satrajit and Nguyen, Phuc and Ghosh, Soumyendu Sekhar and Sotiras, Aristeidis}},
  title   = {{QCResUNet: Joint subject-level and voxel-level segmentation quality prediction}},
  journal = {{Medical Image Analysis}},
  year    = {{2025}},
  doi     = {10.1016/j.media.2025.103718}
}

@article{Jebril_2025_IEEE_Access,
  author  = {{Jebril, Haneen and Pinetz, Thomas and Bogunović, Haris}},
  title   = {{Shape Prior For Quality Assessment in OCTA via Denoosing Autoencoders at the Segmentation Level}},
  journal = {{IEEE Access}},
  year    = {{2025}},
}

@book{KITS2021,
  title={Kidney and Kidney Tumor Segmentation},
  subtitle={MICCAI 2021 Challenge, KiTS 2021, Held in Conjunction with MICCAI 2021, Strasbourg, France, September 27, 2021, Proceedings},
  editor={Nicholas Heller and Fabian Isensee and Darya Trofimova and Resha Tejpaul and Nikolaos Papanikolopoulos and Christopher Weight},
  year={2022},
}

@article{MedSAM,
  title={Segment anything in medical images},
  author={Ma, Jun and He, Yuting and Li, Feifei and Han, Lin and You, Chenyu and Wang, Bo},
  journal={Nature Communications},
  volume={{15}},
  number={{1}},
  pages={{654}},
  year={{2024}},
  doi={10.1038/s41467-024-44824-z}
}

@article{fournel2021medical,
  title={{Medical image segmentation automatic quality control: A multi-dimensional approach}},
  author={Fournel, Joris and Bartoli, Axel and Bendahan, David and Guye, Maxime and Bernard, Monique and Rauseo, Elisa and Khanji, Mohammed Y and Petersen, Steffen E and Jacquier, Alexis and Ghattas, Badih},
  journal={Medical Image Analysis},
  year={2021},
}

@article{rebain2022attention,
  title={{Attention beats concatenation for conditioning neural fields}},
  author={Rebain, Daniel and Matthews, Mark J and Yi, Kwang Moo and Sharma, Gopal and Lagun, Dmitry and Tagliasacchi, Andrea},
  journal={arXiv preprint arXiv:2209.10684},
  year={2022}
}

@article{
doi:10.1073/pnas.2216399120,
author = {Benjamin Billot  and Colin Magdamo  and You Cheng  and Steven E. Arnold  and Sudeshna Das  and Juan Eugenio Iglesias },
title = {Robust machine learning segmentation for large-scale analysis of heterogeneous clinical brain MRI datasets},
journal = {Proceedings of the National Academy of Sciences},
year = {2023},
}

@inproceedings{Higgins2016betaVAELB,
  title={beta-VAE: Learning Basic Visual Concepts with a Constrained Variational Framework},
  author={Irina Higgins and Lo{\"i}c Matthey and Arka Pal and Christopher P. Burgess and Xavier Glorot and Matthew M. Botvinick and Shakir Mohamed and Alexander Lerchner},
  booktitle={International Conference on Learning Representations},
  year={2016}
}

@InProceedings{robinson2018realtime,
author="Robinson, Robert
and Oktay, Ozan
and Bai, Wenjia
and Valindria, Vanya V.
and Sanghvi, Mihir M.
and Aung, Nay
and Paiva, Jos{\'e} M.
and Zemrak, Filip
and Fung, Kenneth
and Lukaschuk, Elena
and Lee, Aaron M.
and Carapella, Valentina
and Kim, Young Jin
and Kainz, Bernhard
and Piechnik, Stefan K.
and Neubauer, Stefan
and Petersen, Steffen E.
and Page, Chris
and Rueckert, Daniel
and Glocker, Ben",
title="Real-Time Prediction of Segmentation Quality",
booktitle="Medical Image Computing and Computer Assisted Intervention -- MICCAI 2018",
year="2018",
}

@inproceedings{kohlberger2012evaluating,
  title={{Evaluating segmentation error without ground truth}},
  author={Kohlberger, Tim and Singh, Vikas and Alvino, Christopher and Bahlmann, Claus and Grady, Leo},
  booktitle={International Conference on Medical Image Computing and Computer-Assisted Intervention},
  year={2012},
}

@article{zhou2021review,
  title={A review of deep learning in medical imaging: Imaging traits, technology trends, case studies with progress highlights, and future promises},
  author={Zhou, S Kevin and Greenspan, Hayit and Davatzikos, Christos and Duncan, James S and Van Ginneken, Bram and Madabhushi, Anant and Prince, Jerry L and Rueckert, Daniel and Summers, Ronald M},
  journal={Proceedings of the IEEE},
  volume={109},
  number={5},
  pages={820--838},
  year={2021},
  publisher={IEEE}
}

@article{CHAOS2021,
    title = {{CHAOS Challenge - combined (CT-MR) healthy abdominal organ segmentation}},
    journal = {Medical Image Analysis},
    year = {2021},
    author = {A. Emre Kavur and N. Sinem Gezer and Mustafa Barış and Sinem Aslan and Pierre-Henri Conze and Vladimir Groza and Duc Duy Pham and Soumick Chatterjee and Philipp Ernst and Savaş Özkan and Bora Baydar and Dmitry Lachinov and Shuo Han and Josef Pauli and Fabian Isensee and Matthias Perkonigg and Rachana Sathish and Ronnie Rajan and Debdoot Sheet and Gurbandurdy Dovletov and Oliver Speck and Andreas Nürnberger and Klaus H. Maier-Hein and Gözde {Bozdağı Akar} and Gözde Ünal and Oğuz Dicle and M. Alper Selver},
  }

@article{Robinson2019Automated,
  title={{Automated quality control in image segmentation: application to the UK Biobank cardiovascular magnetic resonance imaging study}},
  author={Robinson, Ross and Valindria, Vanessa V. and Bai, Wenjia and et al.},
  journal={Journal of Cardiovascular Magnetic Resonance},
  year={2019}
}

@inproceedings{radford2021learning,
  title={Learning transferable visual models from natural language supervision},
  author={Radford, A. and others},
  booktitle={International Conference on Machine Learning (ICML)},
  pages={8748--8763},
  year={2021},
  organization={PMLR}
}

@inproceedings{wang2022medclip,
  title={{MedCLIP}: Contrastive learning from unpaired medical images and text},
  author={Wang, Z. and others},
  booktitle={Proceedings of the 2022 Conference on Empirical Methods in Natural Language Processing (EMNLP)},
  pages={3876--3887},
  year={2022}
}

\end{document}